\pdfoutput=1
\documentclass[aps,prd,twocolumn,superscriptaddress,tightenlines,nofootinbib]{revtex4}
\usepackage{graphicx}
\usepackage{epsfig}
\usepackage{bm}
\usepackage{latexsym,amssymb,amsmath,amsfonts,amssymb,txfonts,pxfonts,wasysym,float}
\usepackage{color}

\newcommand{\beq}[1]{\begin{equation}\label{#1}}
\newcommand{\eeq}{\end{equation}}
\newcommand{\bea}[1]{\begin{eqnarray} \label{#1}}
\newcommand{\eea}{\end{eqnarray}}
\newcommand{\ba}{\begin{array}}
\newcommand{\ea}{\end{array}}

\newcommand{\rf}[1]{(\ref{#1})}

\def\be{\begin{equation}}
\def\ee{\end{equation}}
\def\gs{\mathrel{
   \rlap{\raise 0.511ex \hbox{$>$}}{\lower 0.511ex \hbox{$\sim$}}}}
\def\ls{\mathrel{
   \rlap{\raise 0.511ex \hbox{$<$}}{\lower 0.511ex \hbox{$\sim$}}}}

\newcommand{\half}{\frac{1}{2}}

\newcommand{\Emudep}{E_\mu^{\rm dep}}

\newcommand{\Edep}{E^{\rm dep}}
\newcommand{\Ehad}{E_{\rm had}}
\newcommand{\fracab}{\frac{a}{b}}
\newcommand{\meany}{\langle y \rangle}

\newcommand{\rarr}{\rightarrow}

\def\ie{i.e.\ }

\newcommand{\nue}{\nu_{e}}
\newcommand{\numu}{{\nu_\mu}}
\newcommand{\nutau}{\nu_{\tau}}

\newcommand{\postscript}[2]{\setlength{\epsfxsize}{#2\hsize}
   \centerline{\epsfbox{#1}}}

\newcommand{\comment}[1]{}

\usepackage[usenames,dvipsnames]{xcolor}
\definecolor{orange}{cmyk}{0,0.5,1,0}
\definecolor{rossoCP3}{cmyk}{0,.88,.77,.40}
\definecolor{graa}{rgb}{0.8,0.8,0.8}
\definecolor{blaa}{rgb}{0.2,0.2,0.6}

\begin{document}

\title{\color{rossoCP3}{Evidence for a break in the spectrum of
    astrophysical neutrinos
}}

\author{Luis A. Anchordoqui}
 \email{luis.anchordoqui@gmail.com}
\affiliation{Department of Physics \& Astronomy,  Lehman College, City University of
  New York, NY 10468, USA}
\affiliation{Department of Physics,
 Graduate Center, City University
  of New York,  NY 10016, USA}
\affiliation{Department of Astrophysics,
 American Museum of Natural History, NY
 10024, USA}
\author {Martin M. Block}
\thanks{Deceased}
\affiliation{Department of Physics \& Astronomy, Northwestern University, Evanston, IL 60208}
\author {Loyal Durand}
\email{ldurand@hep.wisc.edu}
\affiliation{Department of Physics, University of Wisconsin, Madison,
  WI 53706, USA}
\thanks{Present address: 415 Pearl Court, Aspen, CO 81611, USA}
\author {Phuoc Ha}
\email{pdha@towson.edu}
\affiliation {Department of Physics, Astronomy, and Geosciences,
  Towson University, Towson, MD 21252, USA} 

\author{Jorge F. Soriano}
\email{jfdezsoriano@gmail.com}
\affiliation{Department of Physics \& Astronomy,  Lehman College, City University of
  New York, NY 10468, USA}
\affiliation{Department of Physics,
 Graduate Center, City University
  of New York,  NY 10016, USA}
  
\author{Thomas J. Weiler}
\email{tom.weiler@vanderbilt.edu}
\affiliation{Department of Physics \& Astronomy, Vanderbilt University, Nashville, TN 37235, USA}

\begin{abstract}
  \vskip 2mm \noindent The announcement by the IceCube Collaboration
  of the observation of 53 astrophysical neutrino candidates in the
  energy range $0.03 \alt E_\nu/{\rm PeV} \alt 2$ has been greeted
  with a great deal of justified excitement. Herein we provide fits of
  single and a broken power-law energy-spectra to these
  high-energy starting events (HESEs). By comparing our  statistical results
  from fits to (background-free) shower HESE data with the spectral shape of muon
  neutrinos recently reported by the IceCube Collaboration, we show
  that there is  ($3 \sigma$) evidence for a break in the spectrum of astrophysical
  neutrinos. After that we use the fitted result to predict the rate
  of Glashow events (in the $\approx 6.3$~PeV region) and {\it
    double-bang} tau neutrino events (in the PeV region) just at the
  threshold of IceCube detection.
  \end{abstract}

\maketitle

\section{Introduction}

In 2012, the IceCube Collaboration reported the observation of two
$\sim 1$~PeV neutrinos, with a $p$-value $2.8\sigma$ beyond the
hypothesis that these events were atmospherically
generated~\cite{Aartsen:2013bka}. The search technique was refined to
extend the neutrino sensitivity to lower
energies~\cite{Schonert:2008is}, resulting in the discovery of an
additional 26 neutrino candidates with energies between 30 TeV and 2
PeV, constituting a $4.1\sigma$ excess for the combined 28 events
compared to expectations from muon and neutrino atmospheric
backgrounds produced by cosmic rays which strike the Earth's
atmosphere~\cite{Aartsen:2013jdh}.  With foresight (and luck) some of
us used these early IceCube data to find the most probable neutrino
spectral index, $\gamma$, assuming a single index describes the data,
with the result $\gamma = 2.3 -
2.4$~\cite{Anchordoqui:2013qsi}. Subsequent studies by the IceCube
Collaboration with a larger data sample bolster our
results~\cite{Aartsen:2014muf}.

At the time of writing, 54 ``high-energy starting events'' (HESE's),
i.e.\ events initiated within the IceCube detector volume by entering
neutrinos, have been reported in four years of IceCube data taking
(1347 days between $2010 - 2014$).  With these events, a purely
atmospheric explanation is rejected at more than
$5.7\sigma$~\cite{Aartsen:2014gkd}.  The data are consistent with
expectations for equal fluxes of all three neutrino
flavors~\cite{Aartsen:2015ivb,Palomares-Ruiz:2015mka}.  The analysis of all four years of
data using an unbroken power law yields a best-fit spectral index of
$\gamma = -2.58 \pm 0.25$, which is compatible with the 3-year
result~\cite{Aartsen:2014gkd}.  While the HESE flux above 200~TeV can
be accommodated by a single power law with a spectral index $\gamma =
2.07 \pm 0.13$~\cite{TeVPA}, lowering the threshold revealed an excess
of events in the $30 - 200$~TeV energy range~\cite{Aartsen:2014muf},
raising the possibilities that the cosmic neutrino spectrum does not
follow a single power law, and/or may be contaminated by an additional
charmed particle background~\cite{Halzen:2016pwl,Halzen:2016thi}.

Indeed, quite recently the IceCube Collaboration reported a combined
analysis based on six different searches for astrophysical
neutrinos~\cite{Aartsen:2015knd}.  Assuming the neutrino flux to be
isotropic and to consist of equal flavors at Earth, the all flavor
spectrum with neutrino energies $25~{\rm TeV} \leq E_\nu \leq 2.8~{\rm
  PeV}$ is well described by an unbroken power law with best-fit
spectral index $-2.50 \pm 0.09$ and a flux at 100~TeV of
$(6.7^{+1.1}_{-1.2}) \times 10^{-18}~({\rm GeV\, s \, sr \,
  cm}^2)^{-1}$. Splitting the data into two sets, one from the
northern sky and one from the souther sky, allows for a satisfactory
power law fit with a different spectral index for each hemisphere.
The best-fit spectral index in the northern sky was found to be
$\gamma_N = 2.0^{+0.3}_{-0.4}$, whereas in the southern sky it was
$\gamma_S = 2.56 \pm 0.12$. The discrepancy with respect to a single
power law is found to be $1.1 \sigma$~\cite{Aartsen:2015knd}.  
It is tempting to speculate
that the different observed spectral indices ($\gamma_N$ and
$\gamma_S$) could be a harbinger of a real anisotropy between the two
hemispheres.  A lower energy contribution to the Southern hemisphere
might be expected since much of the Galactic Plane (including its
center~\cite{Razzaque:2013uoa}) lies in the Southern
hemisphere~\cite{Neronov:2015osa}.  An excess of lower energy events
would push the spectral index of the single power-law Southern
hemisphere to a larger $| \gamma_S |$ value.  The hard spectral index
$\gamma_N$ is supported by a complimentary study using charged current
muon neutrino events where the interaction vertex can be outside the
detector volume~\cite{Aartsen:2016xlq}. This analysis, which includes
IceCube data from 2009 through 2015 with the field of view restricted
to the Northern hemisphere so that the Earth filters out atmospheric
muons, suggests a neutrino spectrum $\propto E_\nu^{-(2.13 \pm
  0.13)}$, for neutrino energies $191~{\rm TeV} \leq E_\nu \leq
8.3~{\rm PeV}$.

Independently of the presence or absence of the Galactic component of
the astrophysical neutrino signal, a significant contribution to the
flux could come from a population of extragalactic cosmic ray sources;
for a review see e.g.~\cite{Anchordoqui:2013dnh}. 

To investigate possibilities, in this article we perform a
  study to constrain the spectral shape of the diffuse neutrino
  flux. Our intent is to establish whether or not a statistically
  significant break exists using data with reasonably well-known
  neutrino energies from 30~TeV to 10~PeV.  This includes essentially
  the entire IceCube range, and as we know, the relative surplus of
  lower energy events and absence of events above 2.3~PeV give
  significant constraints on the spectrum. 
  It is not our intent in this paper to provide an 
  explanation for a break, if it exists.
  See~\cite{He:2013zpa,Chen:2014gxa,Anchordoqui:2015lqa,Chianese:2016kpu}
  for some illustrative examples of two-component models.

  The layout of the paper is as follows.  In
  Sec.~\ref{sec:nu_interactions} we provide an overview of neutrino
  detection at IceCube, and describe the different event topologies
  resulting from the universal neutral current (NC) and individual
  charged current (CC) interactions of the three neutrino flavors.  In
  Sec.~\ref{sec:likelihood} we describe the particulars of our
  likelihood approach and present spectral fits to the neutrino
  data. We display results from the analysis of HESE events initiated
  by electron and tau neutrinos, considering single and double
  exponential models.  We also show a fit to the entire HESE data
  sample to ascertain whether the event topologies characteristics of
  muon showers are consistent with the fit including all particle
  species. As discussed below, knowledge of the incident muon neutrino
  energy $E_{\nu}$ does not predict the energy deposited in the
  IceCube detector by the muon track $\Emudep$, and vice versa; the
  statistical relation between the two is derived in
  Appendix~\ref{sec:App}. Our analysis is similar in spirit and
  procedure to that in~\cite{Vincent:2016nut}.  However, a key
  difference is that we consider only shower events, so that we have a
  sample of events free from atmospheric-background (even the
  $\numu$~NC is not expected to give background events in our small
  sample).  The trade off in lowered statistics is more than
  compensated by the purity of the sample events.  Then, by comparing
  our shower-only analysis with the recent IceCube study on muon
  events, we are able to establish for the first time a break in the
  spectrum at the $3\sigma$ (99.7\% CL) level.  This evidence for the
  broken power law is our main result.  The prospects for the not so
  distant future, including our predictions for the Glashow
  events~\cite{Glashow:1960zz} and {\it double-bang} tau neutrino
  events~\cite{Learned:1994wg} from the Southern and Northern skies
  are presented in Sec.~\ref{sec:GlashBangs}.  The paper wraps up with
  some conclusions presented in Sec.~\ref{sec:conclusions}.

\section{Neutrino interactions at IceCube}
\label{sec:nu_interactions}

Neutrino (antineutrino) interactions in the Antarctic ice sheet can be
reduced to three categories: {\it (i)}~In CC interactions the neutrino
becomes a charged lepton through the exchange of a $W^{\pm}$ with some
nucleon $N$, $\nu_\ell (\bar \nu_\ell) + N \to \ell^\pm + {\rm
  anything}$, where lepton flavor is labeled as $\ell
\in\{e,\mu,\tau\}$.  {\it (ii)}~In NC interactions the neutrino
interacts via a $Z$ transferring momentum to jets of hadrons, but
producing a neutrino rather than a $\ell^\pm$ in the final state:
$\nu_\ell (\bar \nu_\ell)+ N \to \nu_\ell (\bar \nu_\ell) + {\rm
  anything}$.  The scattered $\nu_\ell$ exits the detector, carrying
away energy, and so the observed energy presents a lower bound for the
incident $\nu_\ell$ energy.  All three neutrino flavors exhibit a NC.
These two possibilities are then projected onto two kinds of IceCube
topologies to yield the three final possibilities: {\it (i)}
``Shower'' (${\cal S}$) events result from all three flavors of NC
events, and from the CC events of the electron and tau neutrinos below
$\sim 2$~PeV.  Shower events (also called ``cascade'' events) refer to
the fact that energy is deposited no charged tracks (produced by muons
or taus) are observed. {\it (ii)}~Below a few PeV, ``track'' (${\cal
  T}$) events are produced only by the muon neutrino CC.  The
$\numu$~CC creates a muon and a hadronic shower within the IceCube
detector, the muon track contributes to the deposited energy, but then
the muon is seen to exit the detector as a single track of unknown
energy.  The deposited energy is only a lower bound to the incident
muon neutrino energy.

At $\nu_\tau$ energies above 3~PeV, $\nu_{\tau}$
CC interactions begin to produce separable {\it double bang}
events~\cite{Learned:1994wg}, with one smaller-energy shower produced by the initial
$\nu_\tau$ collision in the ice, and the second larger-energy shower resulting from
the subsequent $\tau$ decay.  At the lower energies of the data to
which we fit, the showers tend to overlap one another and so are not
discernible; at the energies of our fits, the $\nue$'s and $\nutau$'s
are virtually indistinguishable (see, however,~\cite{Li:2016kra}). The
correlations between the (NC, CC) $\otimes$ (${\cal S}$, ${\cal T}$)
are shown in Table~\ref{tab:topology}.

The classification of observed events in different topologies
  is not always straightforward. While almost all NC $\nu_\mu$ events
  are generally correctly classified as showers, a non negligible
  number of CC $\nu_\mu$ events, of both atmospheric and astrophysical
  origin, could be misclassified as showers if the muon has too
  little energy or is produced near the edge of the detector, escaping
  in both cases without enough energy deposited to be
  detected~\cite{Aartsen:2014muf,Aartsen:2015ivb}.  The effects of
  these misclassifications have been studied in great detail in
  Ref.~\cite{Palomares-Ruiz:2015mka,Vincent:2016nut}.  While
  accounting for misclassifications increases the fraction of
  $\mu$-neutrinos and may have influence on the flavor ratios, with
  present statistics it does not influence the shape of the spectrum
  for a shower plus track analysis~\cite{Palomares-Ruiz:2015mka}. We expect only small differences in
  the cases where only showers are analyzed. In light of this, we
  assume here the event topologies at face value as given
  in~\cite{Aartsen:2014gkd}.

\begin{table}[h!]
\caption{Event topology for each neutrino flavor.}
\begin{tabular}{||c||c|c|c||} 
\hline
\hline
~~~Interaction type~~~& ~~~~~~$e$~~~~~~ & ~~~~~~$\mu$~~~~~~ &
~~~~~~$\tau$~~~~~~ \\\hline \hline CC &${\mathcal S}$&${\mathcal
  T}$&${\mathcal S}$\\ 
\hline
NC &${\mathcal S}$&${\mathcal
  S}$&${\mathcal S}$\\ \hline \hline\end{tabular}
  \label{tab:topology}
  \end{table}

\begin{figure*}
\postscript{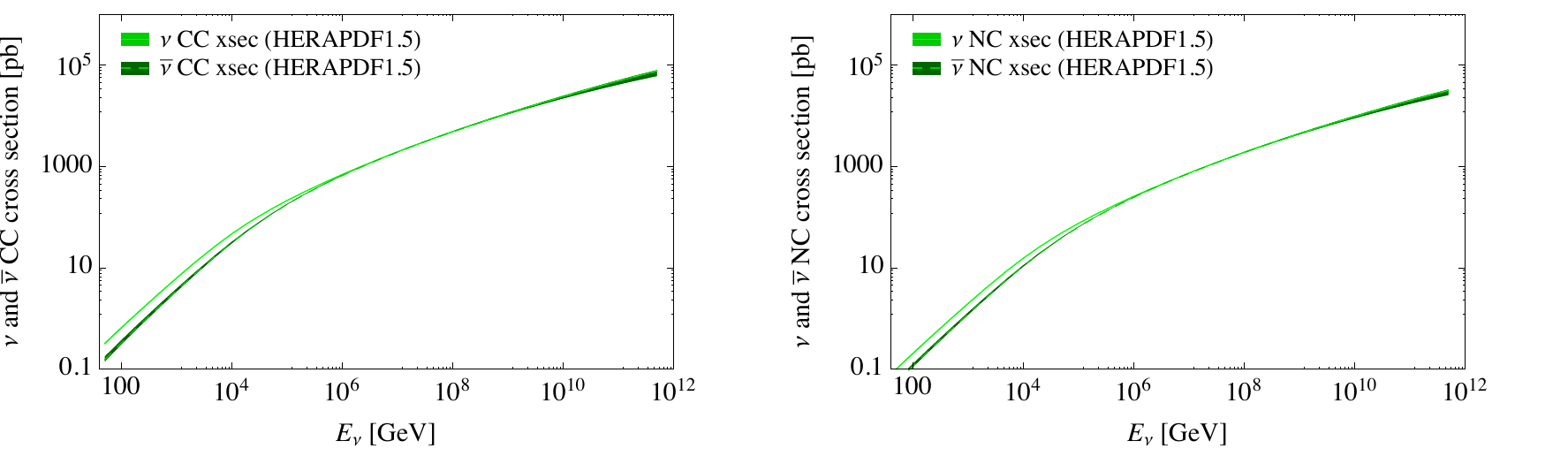}{0.99}
\caption{Neutrino and anti-neutrino cross-sections on isoscalar
  targets for CC and NC scattering according to HERAPDF1.5;
  $\sigma_{{\rm CC}}$ and $\sigma_{{\rm NC}}$, respectively. Taken from
Ref.~\cite{CooperSarkar:2007cv}.}
\label{fig:pQCD}
\end{figure*}

It is appropriate to compare the NC shower rate to the CC shower
  rate. For the reference SM cross sections, we choose the
  results from perturbative QCD calculations constrained by HERAPDF1.5
  shown in Fig.~\ref{fig:pQCD}. These cross sections have been the
  benchmarks adopted by the IceCube
  Collaboration~\cite{Aartsen:2013jdh}.  In the SM, over the energy range we explore here, 
  the NC cross section is 29\% of the total cross section, and the CC cross section
  makes up the remaining 71\%.  Moreover, for the NC, the deposited
  shower energy in the SM is 25\% of the incident neutrino energy on
  average, whereas for the CC, the deposited shower energy is 100\% of
  the incident neutrino energy~\cite{Gandhi:1995tf}.  For an energy falling as power law
  with index $\gamma$, the ratio of NC to CC showers at fixed $\Edep$
  is therefore NC/CC$=(\frac{3}{2})(\frac{29}{71})\,(0.25)^\gamma$,
  where the $\frac{3}{2}$ is due to all three flavors contributing to
  the NC showers, but just two flavors contributing to the CC showers.
  This ratio is smaller than 4\% for $\gamma \ge 2$.  In what follows,
  we account for the NC contribution by weighting the IceCube target mass
  with the cross sections shown in Fig.~\ref{fig:pQCD},  and accept the few percent
  under/over estimate of the flux normalization resulting from uncertainties in the weightings.

  We have the three categories of events at this point, CC and NC
  showers and CC tracks. For the CC HESE shower events, no track
  leaves the detector and $\Edep$ equals the incident neutrino energy,
  $E_\nu$.  For the CC HESE track events, some energy leaves the
  detector in the muon track, and a statistical equation relates
  observed $\Edep$ to $E_\nu$, as given in  Appendix~\ref{sec:App}.  
  Moreover, the muon neutrino events are
  plagued by atmospheric backgrounds (mainly at the lower energies).
  It is estimated that in the four years of data collection, for $E_\nu
  \agt   30~{\rm TeV}$, $12.6\pm 5.1$ events are atmospherically-produced
  down-coming (Southern) background muons, and another
  $9.0^{+8.0}_{-2.2}$ events are atmospherically-produced neutrino
  events~\cite{Aartsen:2014gkd}. The ratio of atmospherically-produced $\numu$'s to $\nue$'s
  is order ten~\cite{Aartsen:2012uu}, and so the atmospheric
  contamination that plagues the non-atmospheric $\numu$ CC is not
  present for our sample of $\nue$ or $\nutau$ CC, or for the NC interactions of all
  flavors.  Accordingly, we choose to analyze just two of the three
  original categories of events, namely the NC and CC shower events, 
  and avoid the track events completely. 
  Later in this paper we will analyze how CC muon track events would
  impact the results of the present shower analysis.

  At the energies of existing data, $\nu_e$'s and $\nu_\tau$'s are
  indistinguishable in their interactions.  The electromagnetic
  cascade triggered by the CC interactions of $\nu_e$ and $\nu_\tau$
  ranges out quickly.  Such a cascade produces a nearly spherical
  light profile, and therefore exhibits a low angular resolution of
  about $15^\circ$ to $20^\circ$~\cite{Aartsen:2013jdh}.  However, a
  fully or mostly contained shower event provides a relatively precise
  measurement of the $\nu_{e/\tau}$ energy, with a resolution of
  $\Delta (\log_{10} E_\nu) \approx 0.26$~\cite{Abbasi:2011ui}.  We
  note that the quality of the energy and angle inference is reversed
  for the CC interactions of $\nu_{\mu}$ induced events.  In this
  case, the secondary muon leaves behind a track of Cherenkov light of
  length a km or more.  Muon tracks point nearly in the direction of
  the original $\nu_{\mu}$, allowing one to infer the arrival
  direction from the observed track with high angular resolution (say
  $\sim 0.7 ^\circ$).  On the other hand, the {\em electromagnetic
    equivalent energy deposited} $\Emudep$ represents only a lower
  bound on the genuine $\nu_{\mu}$ energy.  The authentic $\nu_\mu$
  energy may be up to a factor 5 larger than the deposited energy.  NC
  interactions of all $\nu$ flavors also produce showers, but with a
  rate  60\% (\ie $\frac{3}{2}\times\frac{29}{71}$)  that of CC interactions at the same incident
  neutrino energy, and with a much smaller shower energy  (\ie $(0.25)^\gamma$) , as we
  have already noted.

\begin{figure}
\postscript{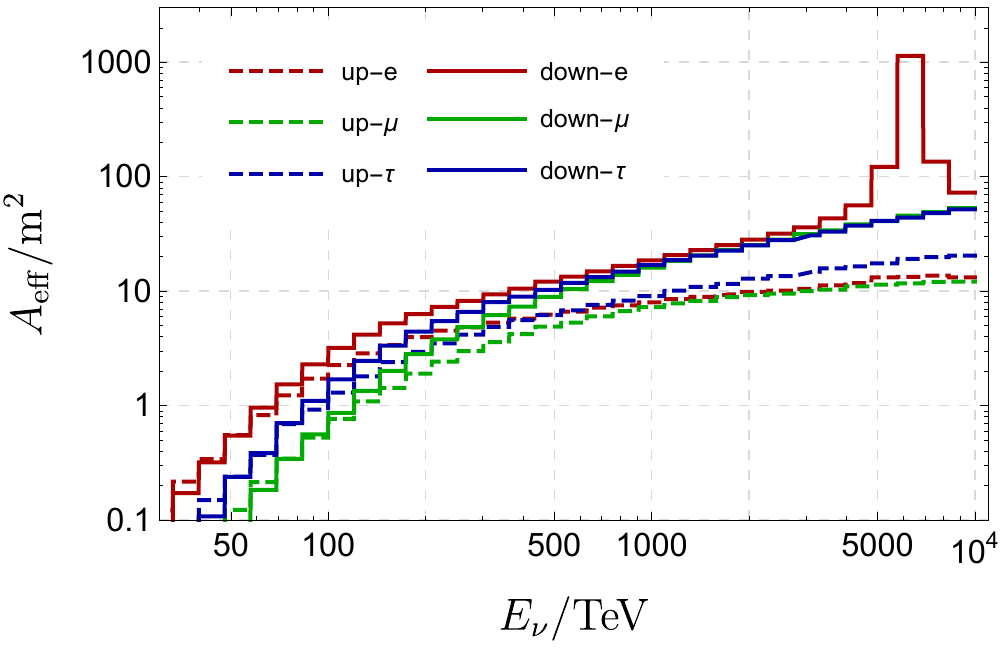}{0.99}
\caption{IceCube effective areas for the different neutrino species.}
\label{Aeff}
\end{figure}

Each of these mentioned effects, and the geometric particulars of the
IceCube detector, are included in the effective areas for HESE events
which have been published by the IceCube
Collaboration~\cite{Aartsen:2013jdh} and are shown in Fig.~\ref{Aeff}.
Conveniently, these $A_{\rm eff}(E_\nu)$'s are separated into those
for the Northern hemisphere (up-going for IceCube at the South Pole)
and those for the Southern hemisphere (down-going for IceCube).
Included in this separation of effective areas is the absorption of
up-going (Northern) neutrinos by the Earth matter.  Thus, the
systematics differ for the Norther and Southern neutrinos, but is
encapsulated in the $A_{\rm eff}(E_\nu)$'s.
(What is not included in the $A_{\rm eff}(E_\nu)$ of $\numu$ is the
relation between $\Emudep$ and $E_{\numu}$, which we provide in  Appendix~\ref{sec:App}.)

\begin{figure}[tpb]
\postscript{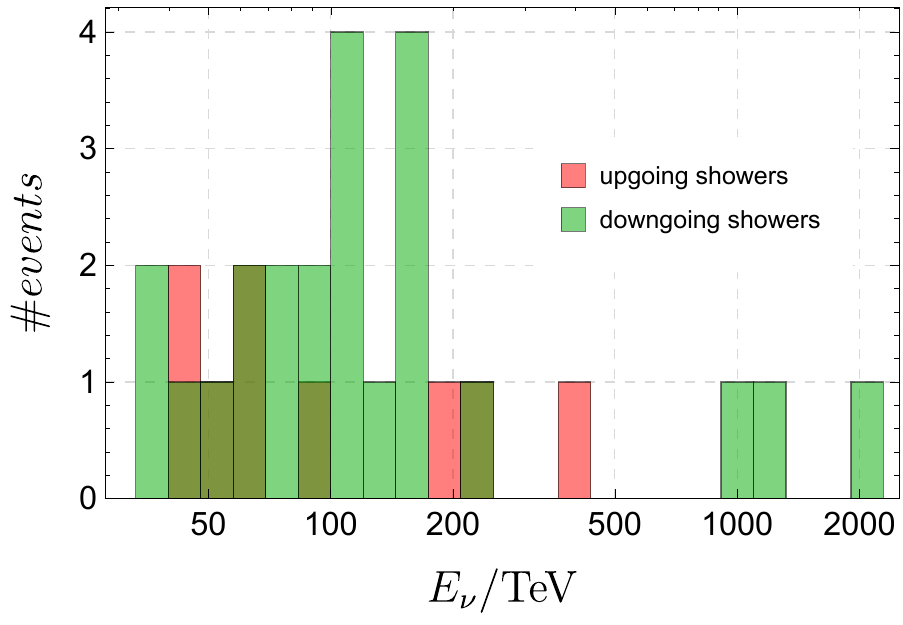}{0.9}
\caption{Number distributions for up- and down-going HESE events.
(Each dark ``event'' is an overlap of two events, one up-going and the other down-going.)}
\label{fig:events}
\end{figure}

In our analysis we use the full 1347-day HESE sample which contains 54
events. One of the events observed in the third year (event \#32) was
produced by a coincident pair of background muons from unrelated
cosmic ray air showers and has now been excluded from the sample.
The remaining events can be classified according to the arrival direction into
North and South. Herein we use the best fit of the arrival direction
to define the North and South subsamples.  Out of the 53 events, 39
are showers. In our analysis we remove the low energy events by
setting an energy threshold $E_\nu \geq 10^{1.52}~{\rm TeV} \simeq
33.11~{\rm TeV}$.  Above this energy there are 32 showers and 14
tracks, the latter events including atmospheric background. 
The numbers for up- and down-going shower events are 9 and 23, respectively.  
The energy distribution of these numbers are  shown in Fig.~\ref{fig:events}.

To summarize this section, for the energy range of present data, there
are two different topologies for the events registered at IceCube,
namely tracks and showers.  The number of track events is expected to
be smaller than the number of shower events by factor of $\sim 6$, and
the background for the track events at lower energies is formidable.
The CC and NC origins of these topologies are summarized in
Table~\ref{tab:topology}.

In the next section we present a full-likelihood approach to fit the
CC and NC shower IceCube data sample, which allows us to constrain the
shape of the energy spectrum of astrophysical neutrinos.

\section{Likelihood analysis}
\label{sec:likelihood}

Armed with IceCube observations we now perform the analysis 
to extract neutrino flux parameters using a maximum likelihood method. For completeness, we first write the most general form for the
likelihood function and then we particularize the study to the different situations of interest.

Let $\boldsymbol\theta$ be the set of
parameters involved in the data analysis, containing all the relevant
guidelines to vary the incident flux.   E.g., the $\boldsymbol\theta$ may be the normalization and spectral index of the power-law fit.
Let $\overline N_{x,k}^{\mathcal Z}$
be the measured number of events with topology ${\mathcal
  Z}\in\{{\mathcal S},{\mathcal T}\}$ and hemispherical direction $x\in\{u,d\}$ in
the energy bin $k$. The probability that the bin $k$ contains $\overline N_{x,k}^{\mathcal Z}$ events of type $(x,{\mathcal Z})$ while
expecting $N_{x,k}^{\mathcal Z}(\boldsymbol\theta)$ is given by a
Poisson distribution \begin{equation} 
{\mathcal P}
\left[
\overline N_{x,k}^{\mathcal Z}
\right.\left| N_{x,k}^{\mathcal Z}(\boldsymbol\theta)
\right]=
\frac{e^{-N_{x,k}^{\mathcal Z}}
 \left(
 N_{x,k}^{\mathcal Z}
 \right)^{\overline N_{x,k}^{\mathcal Z}}}
{{\overline N_{x,k}^{\mathcal Z}! }}\,,
\label{eq:poiss}
\end{equation} 
while the probability that
the bin $k$ contains $\overline N_{x,k}^{\mathcal Z}$ events of type
$(x,{\mathcal Z})$ for all the types is 
\begin{equation}\mathcal
  P_k(\boldsymbol\theta)\equiv\prod_{x,{\mathcal Z}}\mathcal P\left[\bar
    N_{x,k}^{\mathcal Z}\right.\left|N_{x,k}^{\mathcal
      Z}(\boldsymbol\theta) \right].
      \end{equation} 
The likelihood of
having a given a set of parameters $\boldsymbol\theta$ observing the
actual event distribution is 
\begin{equation}
\mathcal  L(\boldsymbol\theta)=\prod_k
\mathcal  P_k(\theta).
  \label{eq:like}
  \end{equation} 
By the maximization of
$\mathcal L$ in terms of the parameters $\boldsymbol\theta$ we will
estimate the most likely values for those parameters.
The logarithm of the likelihood is often taken to ensure that we work with sums instead of with products.
Thus, as an alternative formulation, 
maximization of $\mathcal L$ becomes minimization of $-\ln \mathcal L(\boldsymbol\theta)$.
We have $-\ln {\cal P}_k = \sum_{x,\mathcal Z} \left[ 
 N_{x,k}^{\mathcal Z}  + {\overline N}_{x,k}^{\mathcal Z}  \ln \left( N_{x,k}^{\mathcal Z}\right)  -\ln \left( {\overline N}_{x,k}^{\mathcal Z}\,! \right)\,\right]$. 
 The latter term, $-\ln \left( {\overline N}_{x,k}^{\mathcal Z}\,! \right)$, may be continued as a Gamma function:
 $-\ln \Gamma \left( {\overline N}_{x,k}^{\mathcal Z} +1 \right) $. 
 Notice that in bins where there are zero events, the log-likelihood still receives a nonzero contribution 
 $ \sum_{x,\mathcal Z} N_{x,k}^{\mathcal Z}$ 
 (The Poisson likelihood for an empty bin is $e^{-\sum_{x,\mathcal Z} N_{x,k}^{\mathcal Z}}$). 

The expected number of events per bin is given by
\begin{equation}
N^{\cal Z}_{x,k} (\theta) = 2 \pi T \, \int_k \Phi_j (E_\nu, \theta) \, A_x^{\cal
  Z} (E_\nu,\theta) \, dE_\nu \,,
\end{equation}
where, $\Phi_j$ is the diffuse neutrino flux per flavor and per
particle/antiparticle, with $j$ taking values in $\{ \nu_e, \bar
\nu_e, \nu_\mu,\bar \nu_\mu, \nu_\tau, \bar \nu_\tau\}$, and
$A_x^{\cal Z}$ is the effective area  in the $k$-th energy bin, and
where $\int_k$ represents the integration along that bin.  
For sufficiently narrow bins in $\ln
E_\nu$, it can easily be shown that the integral is well approximated
by 
\begin{eqnarray}
N_{x,k}^{\mathcal Z}(\boldsymbol\theta) & = & 2\pi T
\left<A_x^{\mathcal Z}\right>_k  \int_k \Phi_j
(E_\nu,\boldsymbol\theta) \, dE_\nu \label{eq:ntotal}
\end{eqnarray} where $\left<A_x^{\mathcal Z}\right>_k$ is the averaged
effective area for $(x,\mathcal Z)$. For local power-law descriptions of the effective area and the flux, the corrections are of order $\Delta_k^2$,
where $\Delta_k$ is the width of bin $k$. These corrections are
negligible for the IceCube bins. The non averaged effective areas for $(x,\mathcal Z)$ events are obtained as 
\begin{equation}
A_x^{\mathcal Z}=\sum_{(i,\ell)\in\mathcal Z} A_x^{i,\ell}.\label{eq:avgareas}
\end{equation}
Here $i$ labels the interaction type (charged or neutral current) and $\ell$ labels the neutrino flavor,  
and sums are extended to the values allowed by Table \ref{tab:topology} for each topology. 
Finally, $A_x^{i,\ell}=\omega ^{i,\ell} A^\ell_x$, being $A_x^\ell$ the effective areas accompanying~\cite{Aartsen:2013jdh}. The weights $\omega^{i,\ell}$ can be calculated from the target-mass data (also in \cite{Aartsen:2013jdh}) as
\begin{equation}
\omega^{i,\ell}=\frac{{\eta_i M_i^\ell}}{\sum_k\eta_k M_k^\ell}\,,
\end{equation}
with $\eta_i\equiv\sigma_i/\sigma_{TOT}$ as given in
Fig.~\ref{fig:pQCD}. 

First we perform an  approach which finesses the inevitable
statistical uncertainty in $\Emudep/E_{\numu}$, by simply omitting the
track events from the data sample.  Assuming equal representations of
the three neutrino flavors in the incident neutrino flux, Monte Carlo
simulations reveal the ratio of (up-going) track events at fixed $\Emudep$ to be
of order 1/6 in IceCube~\cite{privateFrancis}.\footnote {A priori one
  would expect the rate to reflect the equal flavor ratios, $\sim
  1/3$, but systematic differences in $\Emudep$ from track and shower
  produce another suppression factor of $\sim 1/2$.  } Thus, the loss
of event statistics due to omission of track events is small, of order
17\%.

\subsection{Unbroken power law}\label{single}

We first hypothesize that the cosmic neutrino flux per flavor and per
particle/antiparticle, averaged over all three flavors, follows an
unbroken power law of the form
\begin{eqnarray}
\Phi_j(E_\nu) \equiv \frac{dF_j}{dE_\nu dA d\Omega dt} &  = & \Phi_0
(E_\nu/E_0)^{-\gamma}  \nonumber \\
& = & \Phi_0 \exp\left[ - \gamma  \ln(E_\nu/E_0)\right]\,,
\label{eq:sp}
\end{eqnarray}
where the normalization energy scale,  $E_0 \simeq 33.11~{\rm TeV}$, is fixed by
the low energy bound of the first used energy bin above $30~{\rm
  TeV}$. 

The single power flux (\ref{eq:sp}) can be integrated to obtain
\begin{equation}
\int_k\Phi_i(E_\nu,\boldsymbol\theta)dE_\nu =\frac{2 \Phi_0 E_0}{\gamma-1}\left(\frac{\left<E\right>_k}{E_0}\right)^{1-\gamma}\sinh \frac{(\gamma-1)\Delta}{2},
\end{equation}
where
\begin{equation}
\Delta\equiv\ln \left(\frac{E_{\mathrm{max}}^k}{E_{\mathrm{min}}^k}
\right) \quad
{\rm and} \quad
\langle E \rangle_k\equiv\sqrt{E_{\mathrm{min}}^kE_{\mathrm{max}}^k}.
\end{equation}
Note that for the logarithmically spaced bins, $\Delta$ is a
constant. For our bin choice, $\Delta=0.08\ln(10) \approx0.18$.

At this stage it is worthwhile to point out that for energies above about 2~PeV the spectral index 
 $\gamma$ must be steeper than~2.4 for $1\sigma$ consistency with the non-observation of Glashow 
$\bar \nu_e + e^- \rarr W^-$ events at 6.3~PeV~\cite{Barger:2014iua}.

\begin{figure}
\postscript{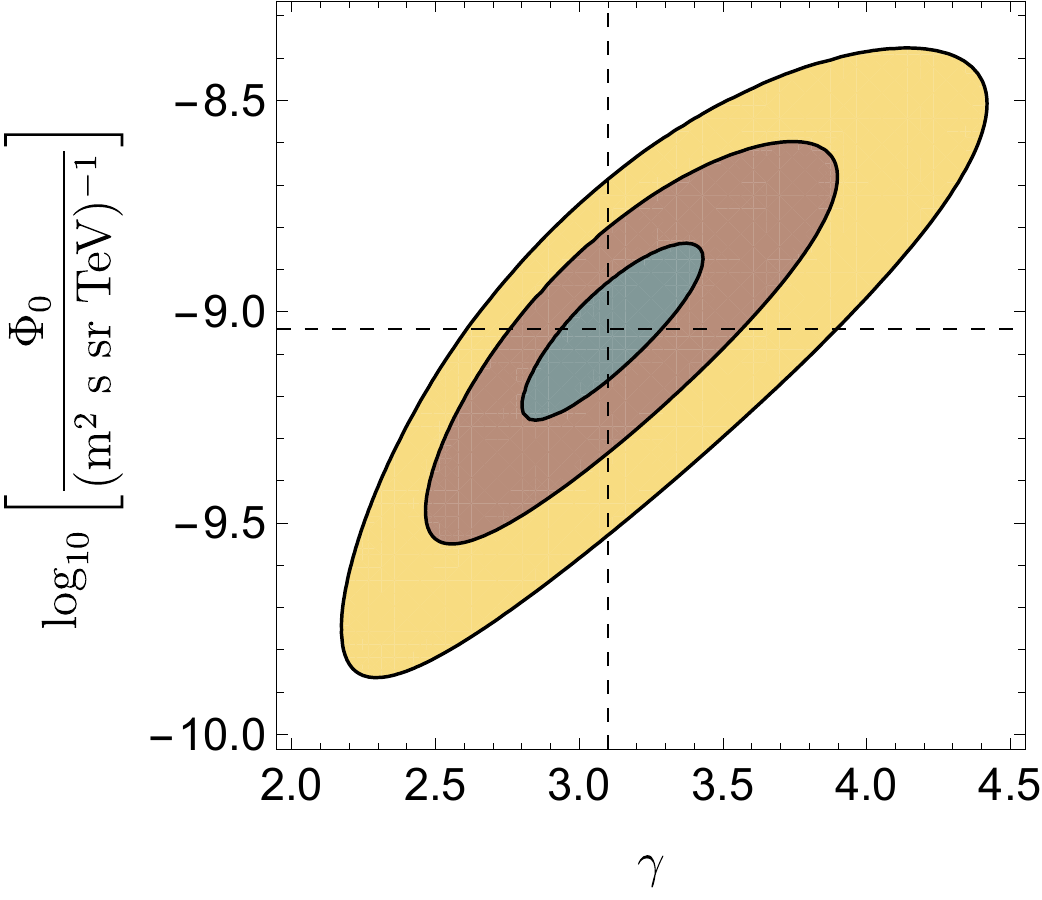}{0.99}
\caption{1, 3, and $5\sigma$ confidence contours for ($\Phi_0, \gamma$).}
\label{comparison}
\end{figure}

\begin{figure}
\postscript{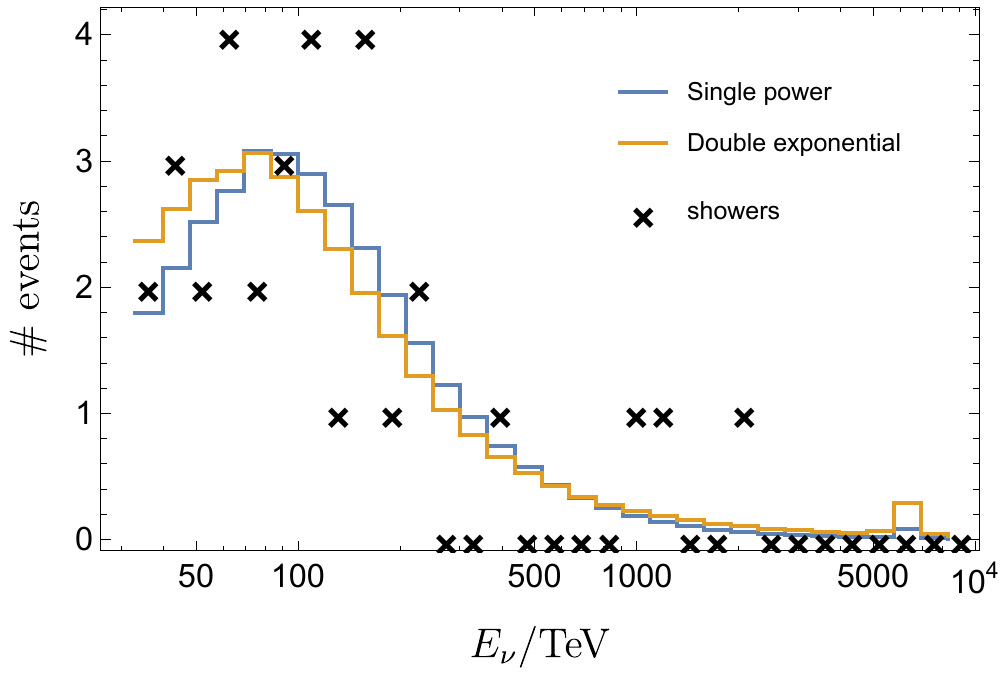}{0.99}
\caption{Histogram of showers, predicted and measured.}
\label{hist_sh}
\end{figure}

\begin{figure}
\postscript{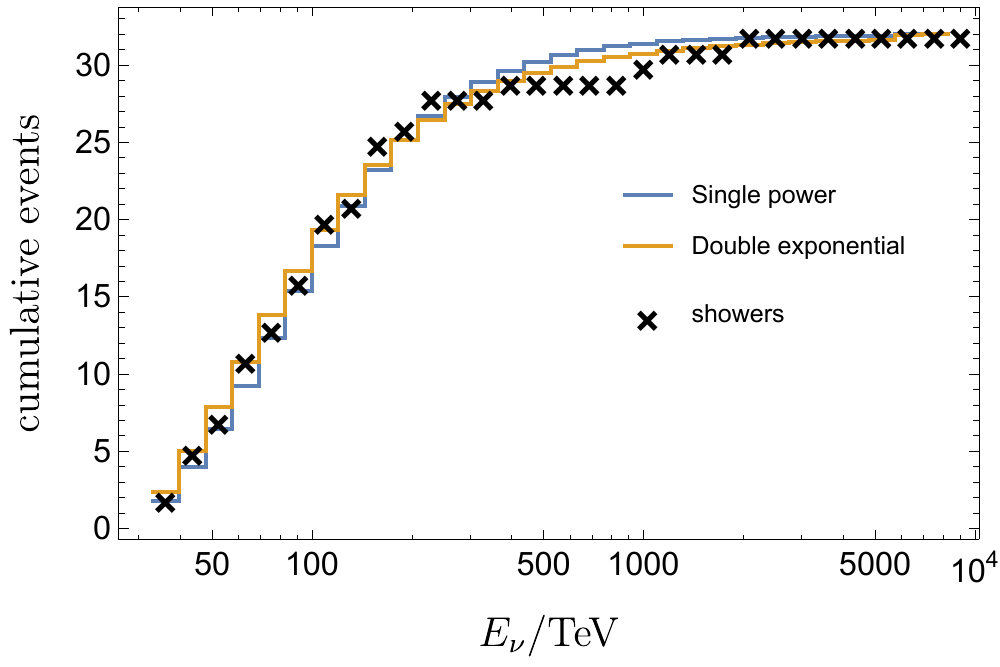}{0.99}
\caption{Cumulative histogram of showers, predicted and measured.}
\label{cum_sh}
\end{figure}

For an unbroken power law, the parameters are
$\boldsymbol\theta=\{\Phi_0,\gamma\}$, for a given reference energy
$E_0$. $\mathcal L(\Phi_0,\gamma)$ can be maximised, using north
and south hemisphere shower data.  The 1, 3, and $5\sigma$ confidence contours are
displayed in Fig.~\ref{comparison}.
After marginalizing the relevant parameters  our results can be
summarized as follows:
\begin{equation}
\Phi_0 =
\left(0.91_{-0.25}^{+0.33}\right)\times10^{-9}\
\mathrm{TeV^{-1}\,m^{-2}\,sr^{-1}s^{-1}} 
\end{equation}
and
\begin{equation}
\gamma = 3.10^{+0.22}_{-0.20} \, .
\label{softgamma}
\end{equation}
In Figs.~\ref{hist_sh} and \ref{cum_sh} we show the local
and cumulative number distributions compared to the data.  We note
that displaying the fit results in this form have some advantages over
the usual log plots of $E_\nu^2 \Phi_j (E_\nu)$ with points and
error bars.  This is because the number distribution is what is
actually measured, and the bins with zero numbers cannot be displayed. 
It is difficult to judge the significance of the results when one's eyes
just follow the flux curve and the non-zero data points. 

On the other hand, we note that plotting of $E_\nu^2\Phi_j (E_\nu)$
versus log $E_\nu$ conserves the area under a spectrum even after
processing the electromagnetic cascade of accompanying $\gamma$
rays. Thus, for optically thin Waxman-Bahcall (WB)
sources~\cite{Waxman:1998yy}, where we expect roughly equal fluxes of photons and
neutrinos, the area of the $\pi^0$ contribution to
the isotropic diffuse $\gamma$-ray spectrum provides an upper bound to
the $\pi^\pm$ origin of the neutrinos.

\begin{figure*}[tbp]
\begin{minipage}[t]{0.49\textwidth}
    \postscript{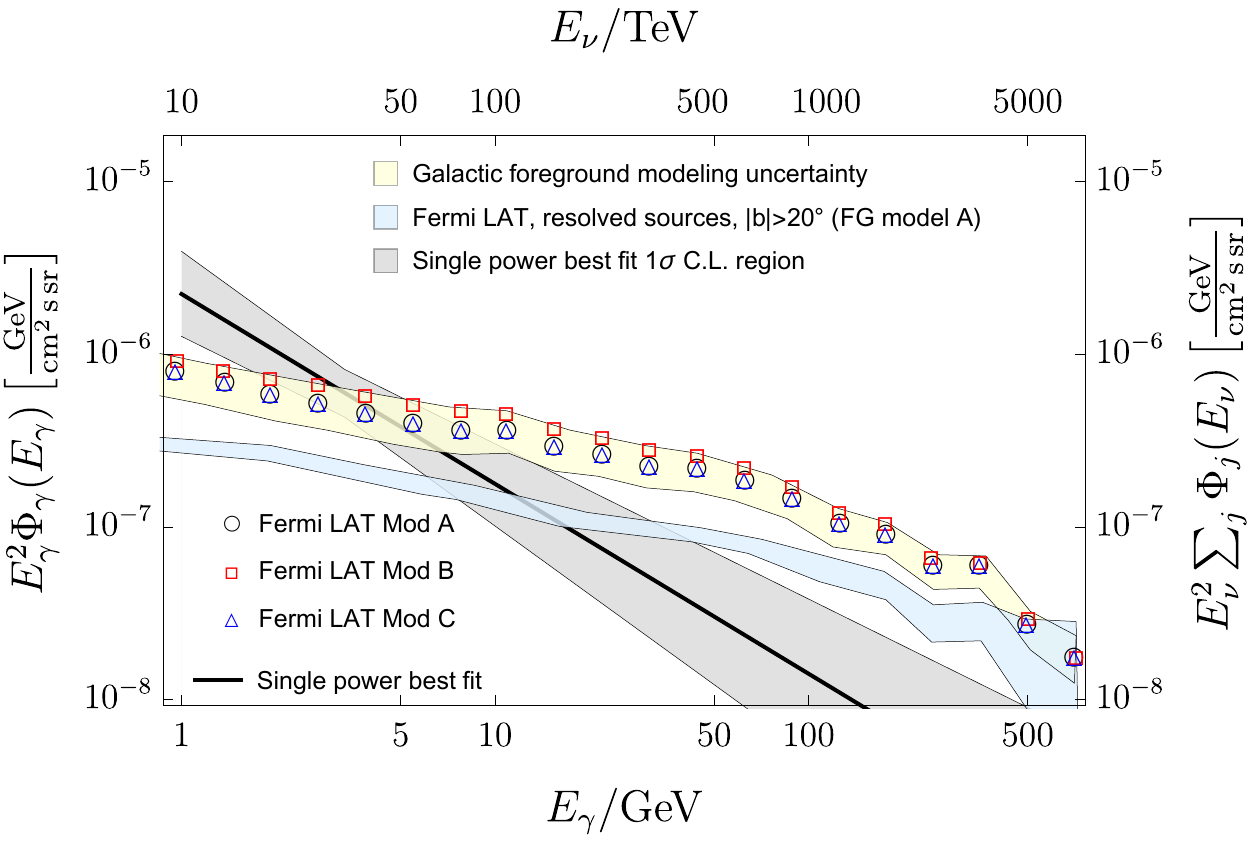}{0.99} 
\end{minipage} 
\hfill
\begin{minipage}[t]{0.49\textwidth}
    \postscript{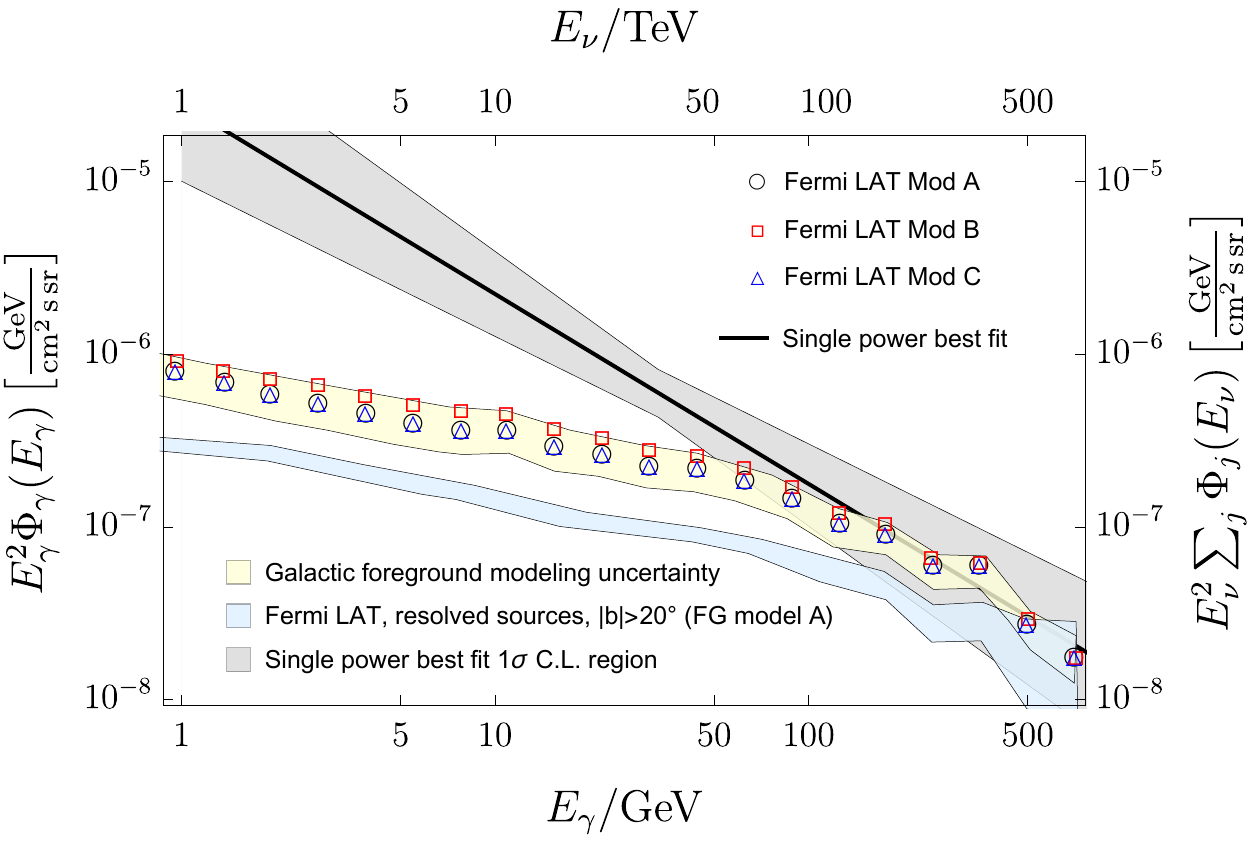}{0.99} 
\end{minipage}
\caption{The open symbols represent the total extragalactic
  $\gamma$-ray background for different foreground (FG) models as
  reported by the Fermi Collaboration~\cite{Ackermann:2014usa}. For details on the modeling of 
the diffuse Galactic foreground emission in the benchmark FG models A,
B and C, see~\cite{Ackermann:2014usa}. The cumulative intensity from
  resolved Fermi LAT sources at latitudes $|b| > 20^\circ$ is indicated by a (grey)
  band.  The best fit to IceCube's shower data (left) and its
  extrapolation down into the TeV-energy range (right) assuming an unbroken power
  law, is also shown
  for comparison. \label{fig:fermi+icecube}}
\end{figure*}

As shown in Fig.~\ref{fig:fermi+icecube} there is tension
between the preferred soft spectral index (\ref{softgamma}) and the
harder spectrum of the isotropic $\gamma$ ray emission measured by
Fermi-LAT~\cite{Ackermann:2014usa}. (The tension explicitly visible in
Fig.~\ref{fig:fermi+icecube} has also been predicted using numerical
simulations~\cite{Murase:2013rfa}.) As of today, this represents the
strongest constraint.  It actually rules out the unbroken power law
hypothesis on the assumption that IceCube's neutrinos are produced via
pion decay in optically thin sources. The tension between
(\ref{softgamma}) and Fermi-LAT data can be relaxed if, for example, a
significant component of the IceCube flux originates in neutron
$\beta$-decay~\cite{Anchordoqui:2003vc}. However, such a possibility
is presently strongly disfavored by the observed neutrino flavor
ratios~\cite{Aartsen:2015ivb}. Alternatively, one can argue that
the extragalactic neutrino sources are hidden, that is, they are opaque to the
emission of $\gamma$ rays and/or cosmic rays producing the neutrino
flux~\cite{Mannheim:1998wp}.  Therefore, it is interesting to
ascertain whether the HESE data by itself imply a break in the
spectrum of astrophysical neutrinos. 

\subsection{Fitting the single power law with showers and tracks}\label{shtr}

To ascertain the impact of $\nu_\mu$ CC interactions in our analysis,
we redo the single power law analysis including the track events
(without subtracting background from tracks). As already noted, the
deposited energy is not the same as the energy of the parent
neutrino. One approximation is to assign an
energy for the parent neutrino of each track event, as explained
in the Appendix~\ref{sec:App}. Otherwise, we proceed exactly as before.
For the single power law model we obtain for the shower plus track events 
\begin{equation}
\Phi_0 =
\left(1.1 \pm 0.3\right)\times10^{-9}~\mathrm{TeV^{-1}\,m^{-2}\,sr^{-1}s^{-1}}
\end{equation}
 and 
\begin{equation}
\gamma = 3.08^{+0.17}_{-0.16} \, .
\label{SBLI}
\end{equation}
Looking at (\ref{softgamma}) and (\ref{SBLI}), we see that our
assumption that misclassification of CC $\nu_\mu$ introduces a
negligible effect is justified when considering only shower
events. This concludes the analysis justifying our assumption that
misclassification of events does not induce significant changes in the
spectral shape.

It is important to stress that for an unbroken power law, spectral
indices $\alt 2.5$ are excluded at 99.7\% CL (from both the shower and
the shower + track analyses). Note, however, that the IceCube analysis
of the muon neutrino spectrum favors an index somewhat harder: 
$\gamma = 2.13 \pm 0.13$~\cite{Aartsen:2016xlq}. This index is $3\sigma$-incompatible 
($2.13 +3\times 0.13=2.52$) with our single power law shower and shower + track analyses, thereby   
providing $3\sigma$~evidence for a break in the spectrum of
astrophysical neutrinos.\footnote{Actually, the statistical
  significance resulting 
  from (\ref{softgamma})  is $2.93\sigma \simeq 3 \sigma$.}

\subsection{Two-exponential model}\label{double}

In this section we describe the broken power law of incident neutrino flux  by the sum of two exponentials in
$\log E_\nu$, or equivalently by two powers of $E_\nu$, 
\begin{equation}
\Phi_j(E_\nu) \equiv  \Phi_0 \left[\left(\frac{E_\nu}{E_0}\right)^{-\gamma_1}+\sigma\left(\frac{E_\nu}{E_0}\right)^{-\gamma_2}\right],
\label{eq:de}
\end{equation}
with $0 < \sigma <1$.

We duplicate our analysis using the maximum likelihood method to extract the values of the
parameters that maximize the probability that the observed number
distributions are described by the assumed flux, for up- and
down-going shower events. The best values of the flux parameters are
found to be:
\begin{equation}
\Phi_0 =\left(1.1_{-0.7}^{+0.6}\right)\times10^{-9}\
\mathrm{TeV^{-1}\,m^{-2}\,sr^{-1}s^{-1}} \,, 
\end{equation}
and
\begin{equation}
\gamma_1 =3.63^{+4.96}_{-0.85}\,, ~~ \gamma_2 = 2.43^{+6.83}_{-0.92}\,,
~~ \sigma =0.0827_{-0.0821}^{+7} \, .
\end{equation}  
The high upper uncertainties observed in the parameters are due to the
flatness of the likelihood function, which comes from the fact that an
increase in one of the parameters can be compensated by decrease in
some of the others to produce an equally likely fit. For example, 
a large value of $\Phi_0$ can be compensated by a value of $\sigma$ low
enough to keep the adequate normalisation of the $\gamma_2$ term, and
with a very large $\gamma_1$, which will make the first term in (\ref{eq:de}) falls rapidly
with increasing energy and not contribute at all to the events over
$33~\mathrm{TeV}$. This basically reduces the fit to a single power-law
with exponent $\gamma_2$ and normalisation $\sigma\,\Phi_0$. Note that
the upper uncertainty of the $\sigma$ parameter  goes into the
unphysical region, $\sigma > 1$. In Figs.~\ref{hist_sh} and \ref{cum_sh} we show the local
and cumulative number distributions for the double exponential model compared to the data.

The significance ($\sim 1.1\sigma$) for the existence of a spectral
break in HESE data is thus comparable to the one obtained by the
IceCube Collaboration splitting a larger data set into northern and
southern subsamples~\cite{Aartsen:2015knd}. It is interesting to note
that the expected number of events above $912~\mathrm{TeV}$ for the
double exponential model is $1.84$, while 3 are observed. The Poisson
probability for this to happen is $0.165$. On the hand, for a single
power law form, the fit predicts that above the same energy $1.013$
events are expected, with an associated Poisson probability of
$0.063$. All in all, for the most likely parameter values, the
double exponential is roughly more probable by a factor of
$0.165/0.063 \approx 2.6$  
than the unbroken power-law.

The break energy is obtained when both flux components are equal. This condition reads
\begin{equation}
E_{\mathrm{break}}=E_0 \ \sigma^{(\gamma_2-\gamma_1)^{-1}}.
\end{equation}
As a practical matter, given the current limited statistics, we
determine the break energy  using the most probable parameter values,
yielding $E_{\rm break} \approx 263~{\rm TeV}$. This break energy is consistent
with the results of~\cite{Vincent:2016nut}.

\begin{figure}
\postscript{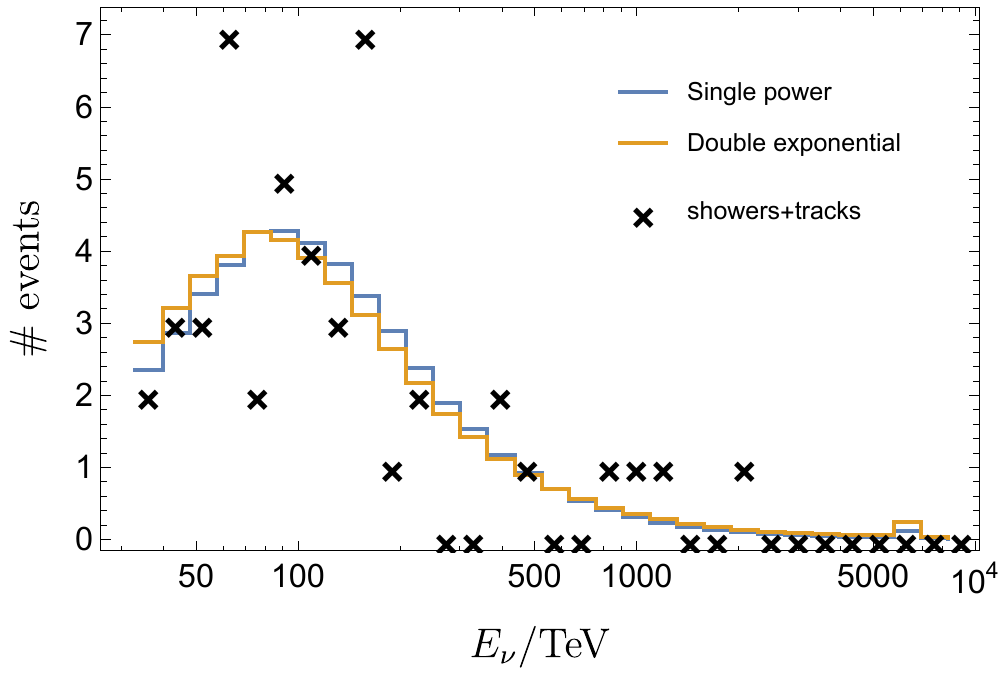}{0.99}
\caption{Histogram of events, predicted and measured.}
\label{hist_shtr}
\end{figure}

\begin{figure}
\postscript{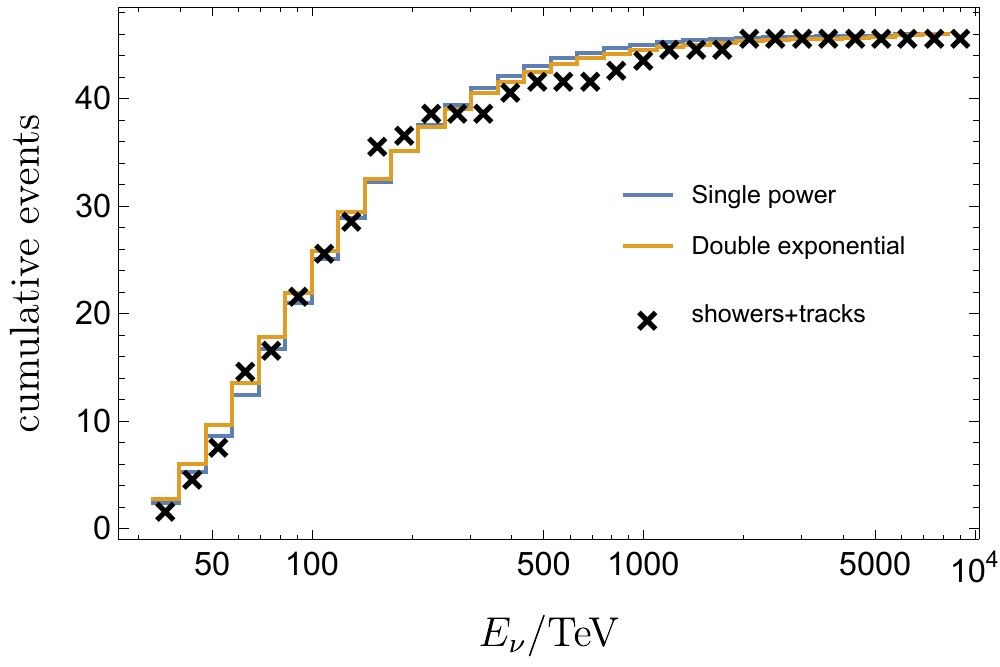}{0.99}
\caption{Cumulative histogram of events, predicted and measured.}
\label{cum_shtr}
\end{figure}

The double exponential model favors the following parameter values
\begin{equation}
\Phi_0 =\left(1.1_{-0.7}^{+0.6}\right)\times10^{-9}\
\mathrm{TeV^{-1}\,m^{-2}\,sr^{-1}s^{-1}} \
\end{equation}
and
\begin{equation} 
\gamma_1 =3.47^{+2.97}_{-0.70} \,, \quad 
 \gamma_2 = 2.62\,, \quad
\sigma =0.17 \, .
\end{equation}  
The results of the likelihood fits can be observed in the histograms and
cumulative histograms shown in Figs. \ref{hist_shtr} and
\ref{cum_shtr}. The break energy is $E_{\rm break} \approx 266~{\rm TeV}$.
It is noteworthy that the confidence intervals of $\gamma_2$ and
$\sigma$ do not close at 68\% C.L. and therefore the result is
consistent with an unbroken power law at the $1\sigma$ level. 

\subsection{Impact of the prompt neutrino flux}

A source of events that could bias our analysis is the
 neutrino flux originated by the prompt decay of charmed particles in the
  atmosphere.  At present, there is a large systematic uncertainty
  in the determination of the prompt neutrino flux. For the lower limit of the allowed prompt neutrino intensity, one would expect no
modification to our previous study.    In order to analyse how a large
  flux of prompt neutrino might alter our results, we repeat the analysis of
  previous sections assuming the other extreme, \ie 
throughout this section we adopt a prompt flux that saturates the upper
     limit for forward charm production ~\cite{Halzen:2016pwl,Halzen:2016thi}. 

\begin{figure}
\postscript{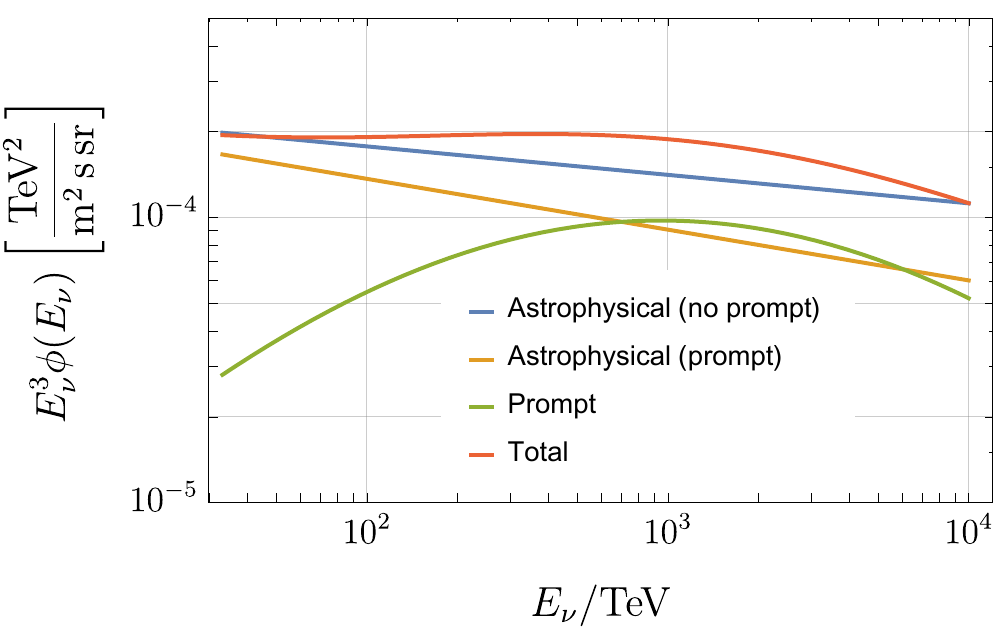}{0.99}
\caption{Prompt, astrophysical and prompt + astrophysical total fluxes for the single-exponential (showers only).}
\label{prompt_se_sh}
\end{figure}

\begin{figure}
\postscript{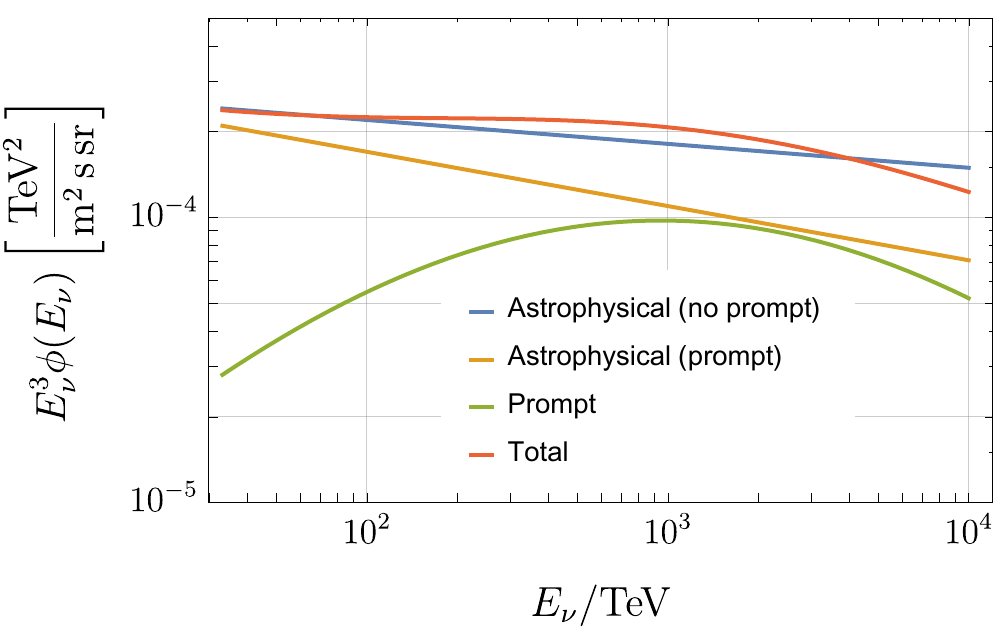}{0.99}
\caption{Prompt, astrophysical and prompt + astrophysical total fluxes for the single-exponential (showers and tracks).}
\label{prompt_se_shtr}
\end{figure}

\begin{figure}
\postscript{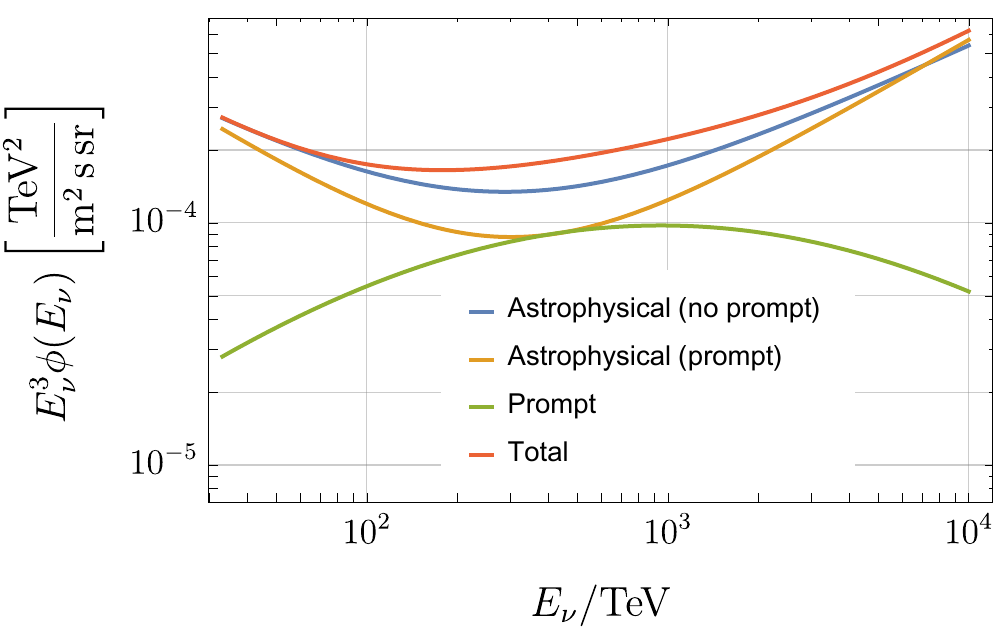}{0.99}
\caption{Prompt, astrophysical and prompt + astrophysical total fluxes for the double-exponential (showers only).}
\label{prompt_de_sh}
\end{figure}

\begin{figure}
\postscript{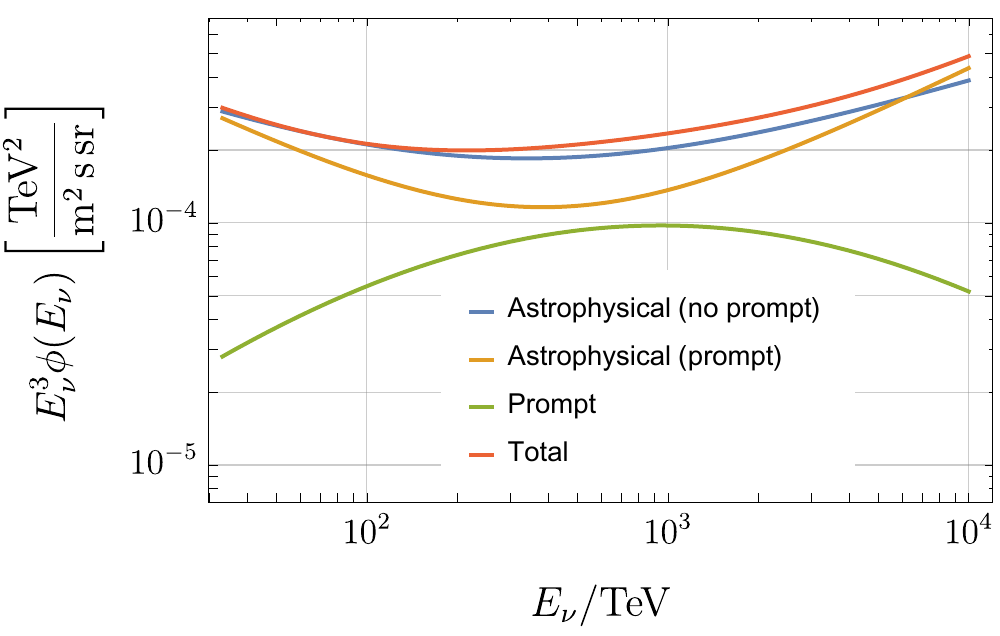}{0.99}
\caption{Prompt, astrophysical and prompt + astrophysical total fluxes for the double-exponential (showers and tracks).}
\label{prompt_de_shtr}
\end{figure}

The fits are done following the same procedure but with slight
modifications in (\ref{eq:ntotal}): {\it (i)} now, the total flux is
the sum of an astrophysical flux component and a parameter-independent prompt flux 
(\ie, it contains no dependence on the fit parameters), 
and {\it (ii)} the flux produced by the charmed particles is
flavor dependent~\cite{Halzen:2016pwl,Halzen:2016thi}. 
The prompt component has
to be corrected to account for the IceCube self-veto applied to the
effective areas as defined in (\ref{eq:avgareas}); this point is
further detailed in Appendix~\ref{sec:App2}. After taking care of
these details, we obtain the results for the fluxes shown in
Figs. \ref{prompt_se_sh}-\ref{prompt_de_shtr}. A comparison with the
results from previous sections (blue lines) is also shown in these figures.

The presence of the prompt component clearly decreases the total
amount of events due to the astrophysical flux, as can be seen both in
the single and double exponential cases. Nevertheless, its energy
dependence favors the power law break in double exponential models, as
can be appreciated in Figs.  \ref{prompt_de_sh} and
\ref{prompt_de_shtr}. The changes on the exponents in both models are
show in Table~\ref{tab:gammas}. The energy break is displaced up to $290\,\mathrm{TeV}$ for showers and $390\,\mathrm{TeV}$ for showers and tracks, approximately.

It is important to emphasize the agreement of our results with those
obtained in~\cite{Halzen:2016pwl,Halzen:2016thi} in the fact that the
prompt flux cannot explain the observed event distribution. In all the
cases the astrophysical flux tends to dominate the spectrum at the
highest energies, and only in the case of single exponential without
tracks the prompt flux dominates the spectrum along some energy range.

\begin{table}[h!]
\caption{Impact of the prompt flux on power laws.}
\begin{tabular}{||c||c|c||} 
\hline
\hline
~~~~Fit model~~~~& ~~~~Without prompt~~~~ & ~~~~With prompt~~~~ \\
\hline \hline
Single, $\mathcal S$ &$3.10$&$3.18$\\ \hline
Single, $\mathcal S+\mathcal T$ &$3.08$&$3.19$\\ \hline
Double, $\mathcal S$ &$(3.63,2.42)$&$(3.78,2.28)$\\ \hline
Double, $\mathcal S+\mathcal T$ &$(3.47,2.62)$&$(3.60,2.38)$\\ \hline \hline\end{tabular}
  \label{tab:gammas}
  \end{table}

\begin{figure}
\postscript{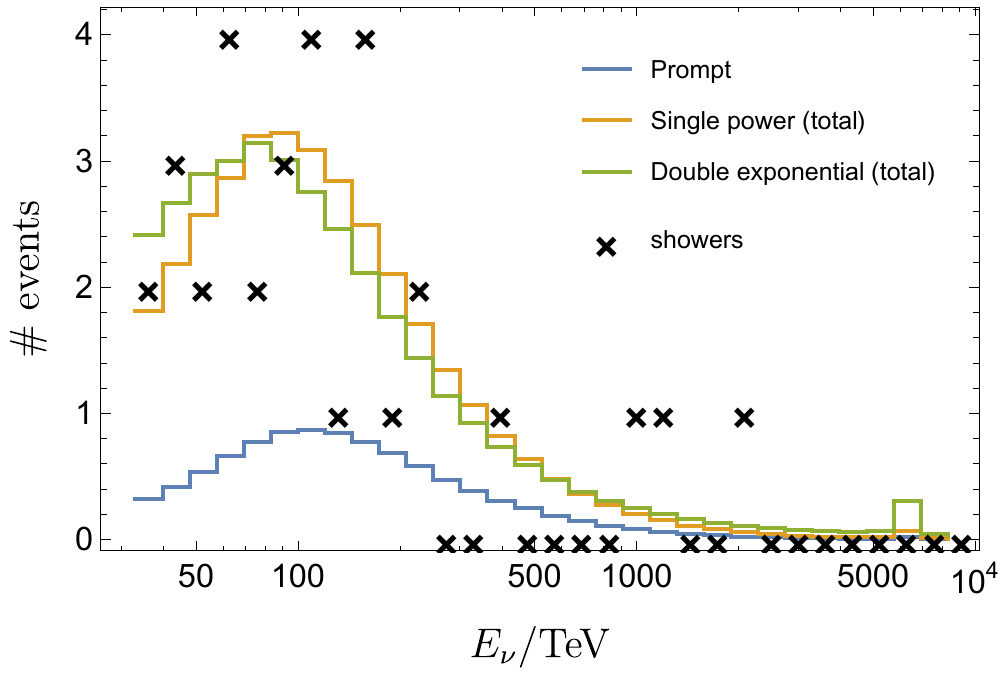}{0.99}
\caption{Histogram of events, predicted and measured, including prompt events (showers).}
\label{hist_sh_prompt}
\end{figure}

\begin{figure}
\postscript{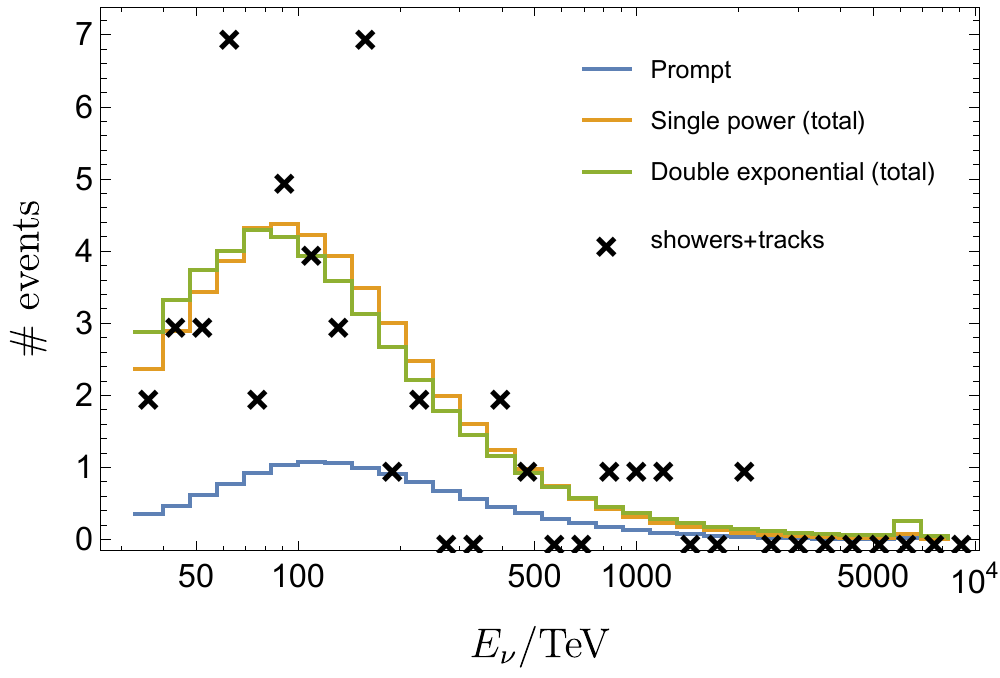}{0.99}
\caption{Histogram of events, predicted and measured, including prompt events (showers and tracks).}
\label{hist_shtr_prompt}
\end{figure}

  Finally, in Figs.~\ref{hist_sh_prompt} and \ref{hist_shtr_prompt} we
  show the final distribution of events. As one can see in Table~\ref{tab:gammas},
  for a single exponential, the effect of the prompt neutrino flux is
  to make the astrophysical neutrino spectrum steeper, and therefore
  our main conclusion from Sec.~\ref{shtr} (there is a
  $3\sigma$ evidence for a break in the spectrum) remains unaltered.

\subsection{Glashow resonometer}

The rate of interaction of $\nu_e$, $\nu_\mu$, $\nu_\tau$, $\bar
\nu_\mu$, $\bar \nu_\tau$ with electrons is mostly negligible compared
to interactions with nucleons. However, the case of $\bar \nu_e$ is
unique because of resonant scattering, 
 $\bar \nu_e +e^+ \to W^- \to$ 
anything, at $E_\nu \simeq 6.3~{\rm PeV}$~\cite{Glashow:1960zz}.  As noted
elsewhere~\cite{Anchordoqui:2004eb}, the signal for $\bar \nu_e$ at the
Glashow resonance, when normalized to the total $\nu + \bar \nu$ flux,
can be used to possibly differentiate between the two primary candidates
($p\gamma$ and $pp$ collisions) for neutrino-producing interactions in
optically thin sources of cosmic rays. In $pp$ collisions, the nearly
isotopically neutral mix of pions will create on decay a neutrino
population with the ratio $N_{\nu_\mu} = N_{\bar \nu_\mu} = 2
N_{\nu_e} = 2 N_{\bar \nu_e}$. On the other hand, in photopion
interactions the isotopically asymmetric process $p \gamma \to
\Delta^+ \to \pi^+n$, $\pi^+ \to \mu^+ \nu_\mu \to e^+ \nu_e \bar
\nu_\mu \nu_\mu$ is the dominant source of neutrinos so that at
production, $N_{\nu_\mu} = N_{\bar \nu_\mu} = N_{\nu_e} \gg
N_{\bar \nu_e}$ (assuming little $\pi^-$ "contamination").  
It is therefore of interest to consider situations in which the flux of
neutrinos is equally divided among the three flavors, but with a
negligible component of $\bar \nu_e$. To account for this possibility 
we consider an effective area which does not contain effects from the Glashow 
resonance. This will allow harder exponents in the high energy component.

The results of the likelihood fit assuming a single exponential model are 
\begin{equation}
\Phi_0 =
\left(1.2_{-0.3}^{+0.4}\right)\times10^{-9}\
\mathrm{TeV^{-1}\,m^{-2}\,sr^{-1}s^{-1}}
\end{equation}
 and 
\begin{equation}
\gamma = 3.17^{+0.22}_{-0.21}
\end{equation}
 for shower events, and 
\begin{equation}
\Phi_0 =
\left(1.4_{-0.3}^{+0.4}\right)\times10^{-9}\
\mathrm{TeV^{-1}\,m^{-2}\,sr^{-1}s^{-1}} \,,
\end{equation}
 with
\begin{equation}
\gamma =3.14^{+0.18}_{-0.17} \,,
\end{equation}
 for showers and tracks together. 
 
\begin{figure}
\postscript{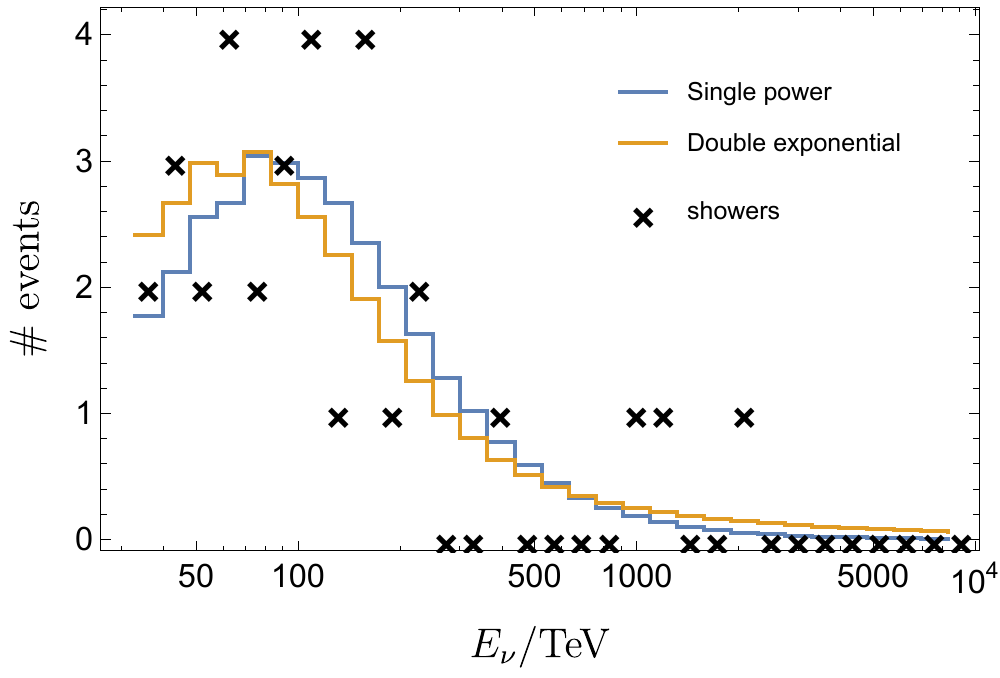}{0.99}
\caption{Histogram of showers, predicted and measured, without Glashow resonance.}
\label{hist_sh_noglashow}
\end{figure}

\begin{figure}
\postscript{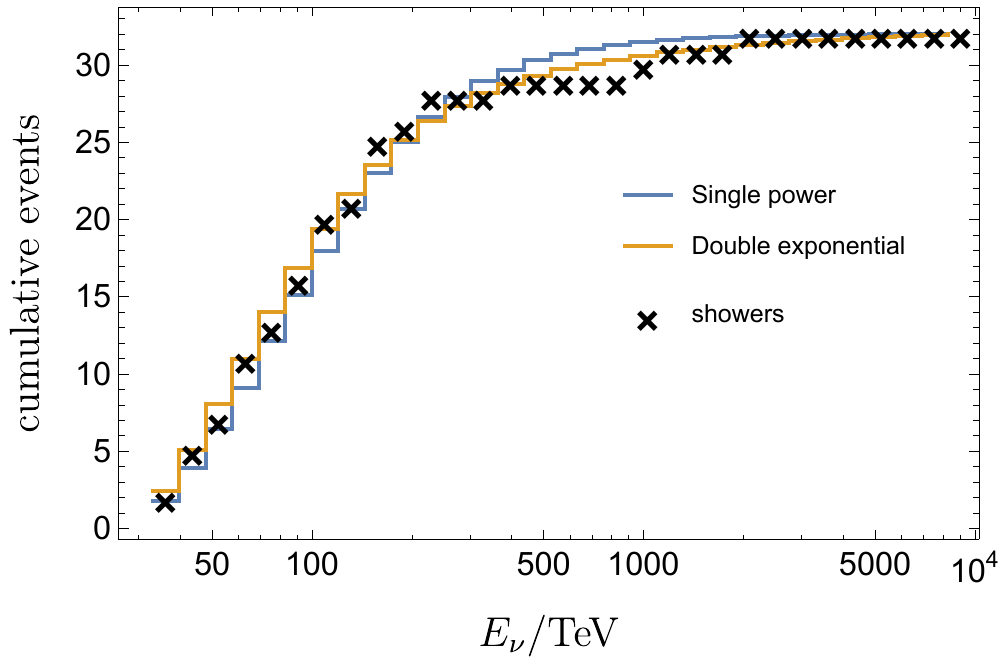}{0.99}
\caption{Cumulative histogram of showers, predicted and measured, without Glashow resonance.}
\label{cum_sh_noglashow}
\end{figure}

\begin{figure}
\postscript{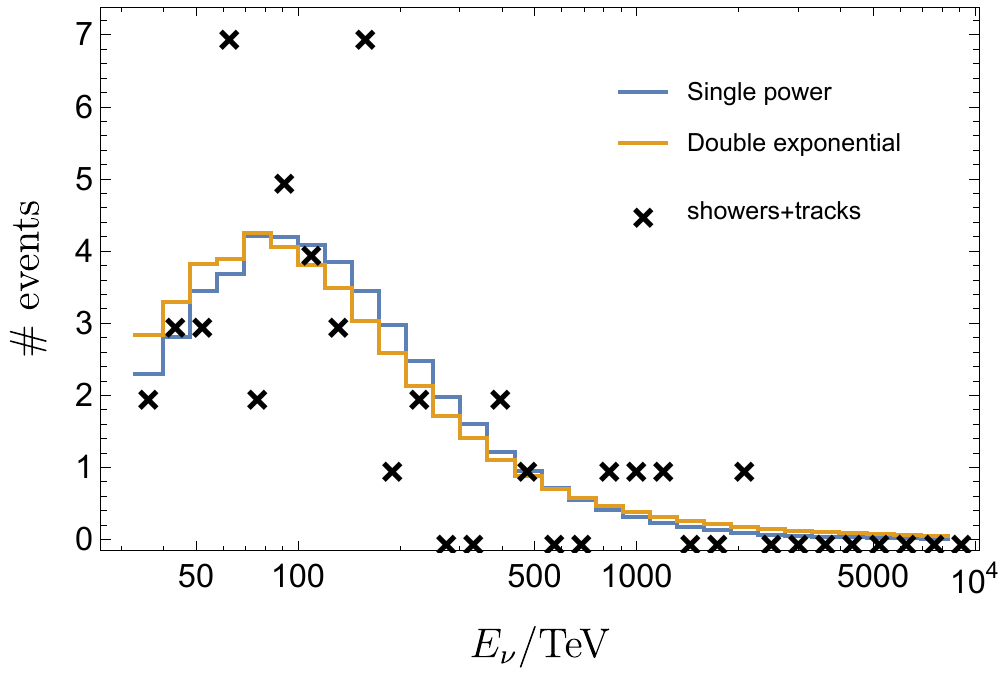}{0.99}
\caption{Histogram of showers and tracks, predicted and measured, without Glashow resonance.}
\label{hist_shtr_noglashow}
\end{figure}

\begin{figure}
\postscript{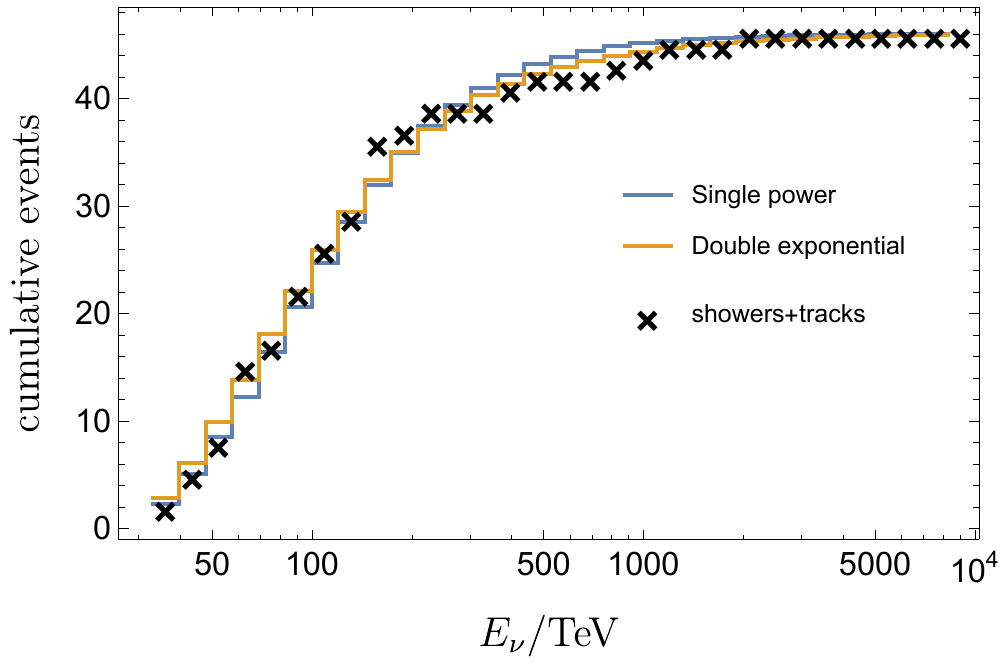}{0.99}
\caption{Cumulative histogram of showers and tracks, predicted and measured, without Glashow resonance..}
\label{cum_shtr_noglashow}
\end{figure}

 For the double exponential model, we obtain
\begin{equation}
\Phi_0 =\left(1.6_{-0.7}^{+0.7}\right)\times10^{-9}\
\mathrm{TeV^{-1}\,m^{-2}\,sr^{-1}s^{-1}},
\end{equation}
and
\begin{equation} 
\gamma_1 =3.63^{+0.90}_{-0.40}\,, ~~ 
\gamma_2 = 2.03^{+0.86}_{-0.84}\,, ~~ 
\sigma =0.0159_{-0.0155}^{+2} \,,
\end{equation}
 for showers, and 
\begin{equation}
\Phi_0 =\left(1.7_{-1.0}^{+0.7}\right)\times10^{-9}\ \mathrm{TeV^{-1}\,m^{-2}\,sr^{-1}s^{-1}}, 
\end{equation}
and
\begin{equation}
\gamma_1 =3.5\,, \quad
\gamma_2 = 2.3\,, \quad
\sigma =0.04 \, ,
\end{equation}
for showers and tracks. 
Note that for showers, 
the upper uncertainty of the $\sigma$ parameter  again goes into the unphysical ($\sigma > 1$) region. 
Moreover,  once more when showers and  tracks are considered in the fit, 
the confidence intervals of $\gamma_1$, $\gamma_2$, and $\sigma$
do not close at 68\%~C.L. and therefore the result is consistent with
a single, unbroken power law at the $1\sigma$ level. The energy break is 
$E_{\rm break} \approx  440~{\rm TeV}$ in the fit to shower data and
$E_{\rm break} \approx 490~{\rm TeV}$ for showers and tracks. The results form the
likelihood fits can be observed in
Figs. \ref{hist_sh_noglashow}, \ref{cum_sh_noglashow},
\ref{hist_shtr_noglashow} and \ref{cum_shtr_noglashow}. 

It is important to stress that the expected number of events above
912~TeV for the double exponential model is 2.167, with a Poisson
probability of 0.194. For the unbroken power law assumption, the fit predicts that above the
same energy 0.899 events are expected, with an associated Poisson
probability of 0.049. Altogether, the double exponential is about
$0.194/0.049 \sim 4$ times more probable than the unbroken power law.

\section{Looking ahead with IceCube-Gen2}

\label{sec:GlashBangs}

In the very near future, two more year's of IceCube data, $2014-2016$,
are expected to be unblinded.  Looking farther into the future, design
studies for the IceCube-Gen2 high-energy array are well
underway~\cite{Aartsen:2014njl}. They will result in an instrumented
volume approaching $10~{\rm km}^3$ and will lead to significantly
larger neutrino detection rates, across all neutrino flavor and
detection channels.\footnote{This may be complemented by atmospheric
  neutrino telescopes~\cite{Neronov:2016zou}.} A rough estimate
indicates about an order of magnitude increase in exposure per
year. The bigger instrumented volume will facilitate the calorimetric
detection of muon tracks, reducing significantly the systematic
uncertainty. The extension will reuse the very reliable design of
IceCube's digital optical modules and therefore it will surely perform
technologically at least at the level of IceCube.  A conservative
estimate of the sample size is then attainable by simply scaling the
instrumented volume.

In 4 years of observation, IceCube has detected 54 events with incident neutrino energy above above 25~TeV.
Of these, about 20 events have energies in the range 100~TeV $ < E_\nu < $2~PeV. 
This detection rate implies that in 10 years of data taken by the IceCube facility will collect on
the order of 50 neutrino events within this (roughly) 
decade of energy. 
The next generation of neutrino telescope in the South pole, IceCube-Gen2, will
increase the per year exposure by about an order of magnitude, and
therefore in 10 year's of observation will collect roughly 500 neutrinos
with 100~TeV $ < E_\nu < $2~PeV.

We have noted that {\it double bang} and Glashow events could play a
key role in constraining processes of neutrino production. It seems
then reasonable to extrapolate the fluxes derived in the previous
section to estimate the event rate at IceCube-Gen2. The search for
{\it double bang} events is possible above 3 PeV. Therefore, we fix 
the search bins by  $ 2.8 < E_\nu/{\rm PeV} <  10$ and determine the
yearly event rate assuming the favored parameters for the double
exponential model. If the neutrino flux is democratically distributed
among flavors and particle-antiparticle, we expect 0.3 {\it double bang} events per year at IceCube-Gen2, 
whereas if the $\bar \nu_e$ component is suppressed, we expect 0.7 {\it double bang} events per year. 
On the other hand, for Glashow events, we search in the resonance bins $4 < E_\nu/{\rm PeV} <  10$ shown in Fig.~\ref{Aeff}.  
As we can see, the numbers of $\nu_\tau$, $\bar\nu_\tau$, and $\bar \nu_e$ events are correlated. 
Therefore, by comparing the rates of Glashow events and {\it double bang} events, 
one will be able to study flavor as well as particle-antiparticle ratios.

\section{Conclusions}
\label{sec:conclusions}

The most pressing consequence of IceCube's discovery of astrophysical
neutrinos is that the flux level observed is exceptionally high by
astronomical standards. The magnitude of the observed flux is at a
level of the WB bound~\cite{Waxman:1998yy} which applies to neutrino
production in optically thin sources that are responsible for
emission of ultrahigh energy cosmic rays. In this paper we have
performed a study to constrain the spectral shape of the diffuse
neutrino flux and obtain information on possible source
environments. Our results are encapsulated in Figs.~\ref{comparison}
to~\ref{cum_shtr_noglashow}, and are summarized in these concluding remarks:

\begin{itemize}

\item We have conducted our study using data from events produced by
  CC interactions of tau and electron neutrino flavors as well as NC
  interactions of all neutrino flavors, but avoiding the background-laden $\numu$~CC track events. 
  The ``shower'' data have essentially no atmospheric background, and and at the same
  time allows a precise determination of the relation between the
  energy deposited in the detector and the original neutrino energy.
  The total atmospheric muon background in four years of data is found to
  be $12.6 \pm 5.1$~\cite{Aartsen:2014gkd}, while the background from
  atmospheric neutrinos reported by the IceCube Collaboration,
  presumed to be overwhelmingly $\nu_\mu$'s, 
  which through their CC produce track topologies~\cite{Aartsen:2012uu},  is
  $9.0^{+8.0}_{-2.2}$~\cite{Aartsen:2014gkd}.

\item If astrophysical neutrinos originate in WB sources, the tension
  between IceCube and Femi-LAT data stands as a strong constraint
  for the hypothesis of an unbroken power law describing the spectrum
  in the energy range $10~{\rm TeV} \alt E_\nu \alt 10~{\rm PeV}$. By
  analyzing IceCube's HESE data sample we have found clues for a break
  in the spectrum at $200 \alt E_{\rm
  break}/{\rm TeV} \alt 500$, independently of the neutrino
  origin(s). Using Poisson statistics we have shown that if the flux
  of neutrinos at Earth is democratically distributed among both
  flavors and particle-antiparticle, then the description of the
  spectrum with an apparent break is roughly 2.6 times more probable
  than the unbroken power law. We have also shown that if the flux of
  $\bar \nu_e$ is significantly suppressed with respect to other
  neutrino species,  so as to suppress resonant production of Glashow events 6.3~PeV, 
  then the description of the spectrum with the
  apparent break is about 4 times more probable than the unbroken
  power law.

\item The IceCube Collaboration has recently released a study using CC
  muon neutrino events, with interaction vertex outside the detector
  volume~\cite{Aartsen:2016xlq}. Because of the large muon range the
  effective area for track topologies is significantly larger than in
  the HESE sample. The price paid is that $\Edep$ is only
  statistically related to a range of incident neutrino energies.  The
  data collected from 2009 through 2015 is well described by a
  Monte-Carlo generated isotropic, unbroken power law flux with a
  normalization at $100~{\rm TeV}$ neutrino energy of
  $(0.90^{+0.30}_{-0.27}􏰁 \times 10^{-18}~ {\rm GeV \ cm^2 \ s \
    sr})^{-1}$ and a spectral index of $\gamma = 2.13 \pm 0.13$.  This
  hard spectrum is consistent with that expected in Fermi
  engines~\cite{Fermi:1949ee}. Our likelihood fit for a neutrino flux
  democratically distributed among flavors and particle-antiparticle,
  yields $\gamma_2 \approx 2.43$; the result of the IceCube
  Collaboration is not inconsistent with our findings.  This is
  because the likelihood function for $\gamma_2$ is rather flat due to
  limited statistics; for all the cases analyzed in this paper, the
  fitted $\gamma_2$ is consistent with a Fermi engine at the 68\%~CL.
  On the other hand, we have shown that for an unbroken power law,
  spectral indices $\alt 2.5$ are excluded at 99.7\%~CL. Requiring the
  IceCube analysis of the $\nu_\mu$ spectral index to be compatible
  with the shower analysis presented here provides additional evidence
  ($3\sigma$ effect) for a break in the spectrum of astrophysical
  neutrinos.

\item A similar study to determine the spectral shape has been
  presented in~\cite{Vincent:2016nut}. However, unlike the study
  of~\cite{Vincent:2016nut}, our statistical analysis becomes
  background free by using only shower events. In such a background
  free analysis, the slope of the spectrum for the unbroken power-law
  hypothesis, $\gamma = 3.10^{+ 0.22}_{-0.20}$, is somewhat steeper
  than the result obtained in~\cite{Vincent:2016nut}, which is $\gamma
  = 2.84^{+0.25}_{-0.27}$.  Our analysis differs
  from~\cite{Vincent:2016nut} in that we obtain smaller errors on the
  neutrino spectral index by sacrificing some data in favor of
  background-free HESE data.  Though the results are compatible at the
  $1\sigma$ level, by direct comparison with the $\nu_\mu$ IceCube
  spectrum ($\gamma = 2.13 \pm 0.13$), we see that only the resulting
  softer shower spectrum and smaller errors resulting from our
analysis allows to extract ($3\sigma$) evidence for a break in the
cosmic neutrino spectrum. For the two-exponential model, the results
of the two studies are also compatible within errors.

\item IceCube has proposed a larger next-generation detector~\cite{Aartsen:2014njl}.
IceCube-Gen2 will surely have a technology at least as sophisticated
as the first generation IceCube, so a conservative estimate of the
future sample size is attainable by simply scaling apertures.
IceCube-Gen2 will have an order of magnitude larger aperture than
IceCube, so one can expect the clarity that comes with at least an
order of magnitude more astrophysical neutrino data. By extrapolating
the flux for a double exponential model we have shown
that if the neutrino flux is democratically distributed among flavors
and particle-antiparticle, then the new South pole facility  will observe about 1 Glashow event per year and
about 0.1  {\it double bangs} per year. If on the other hand the flux of $\bar \nu_e$ is
significantly suppressed with respect to the other neutrino species, then
the rate of {\it double bangs} becomes about 0.7 events per year. Thus by
comparing the rates of Glashow and {\it double bang} events one can study
not only flavor ratios, but also particle-antiparticle ratios. Such analyses
are key to understanding source properties.  

\end{itemize}

In summary, by confronting the favored parameters of our likelihood
fit to shower events assuming an unbroken power law with the hard
$\nu_\mu$ spectrum recently announced by the IceCube Collaboration, we
have shown that there is evidence for a break in the spectrum of
astrophysical neutrinos, sustained by a $3 \sigma$ discrepancy among
the predicted spectral indices by these two analyses.  This is our
main conclusion.  The localization of the break-energy is at present
hampered by the limited available statistics.  However, we have shown
that the favored parameters of our likelihood fit assuming a double
exponential model provide a favored break in the $200 \alt E_{\rm
  break}/{\rm TeV} \alt 500$~range.

\acknowledgments{We thank Francis Halzen for guidance throughout much
  of this work, and Subir Sarkar for permission to reproduce
  Fig.~\ref{fig:pQCD}.  M.M.B., L.D., and T.J.W.\ would like to thank
  the Aspen Center for Physics for its hospitality and for its partial
  support of this work under NSF Grant No. 1066293.  L.A.A.\ is
  supported by NSF Grant No.~PHY-1620661 and by NASA Grant
  No. NNX13AH52G.  P.H.\ would like to thank the Towson University
  Fisher College of Science and Mathematics for support. The research
  of T.J.W.\ was supported in part by DoE Grant No. DE-SC0011981.
  J.F.S.\ was partially supported by the grant Ayudas de movilidad del
  personal docente e investigador 2015, from Universidad de Alcal\'a;
  he thanks Mar\'{\i}a D. Rodr\'iguez Fr\'ias and Luis del Peral for
  their support needed to start this work.  }

\appendix

\onecolumngrid

\section{Energy-dependent Muon Absorption}
\label{sec:App}

As explained in Sec.~\ref{sec:nu_interactions}, at the energies of present IceCube HESE data, NC and CC interactions of the neutrinos 
deposit all their energy into shower energies, except for the CC interaction of the muon neutrino.
For HESE $\numu$ events, a track begins in the IceCube detector, but usually continues beyond the detector's border.
We wish to know the relation between the incident neutrino energy $E_\nu$ and the energy  deposited $\Emudep$ 
in the IceCube detector.  To this end, we begin with the differential equation for the muon energy loss in a medium.
\beq{Emuon loss}
\frac{dE_\mu}{d\ell} = -(a+b\,E_\mu)\,,
\eeq
where $a$ and $b$ are slowly varying functions of muon energy $E_\mu$ 
that also depend on the medium in which the muon propagates.
The coefficient $a$ characterizes the ionization losses of the muon, and $b$ characterizes 
the other losses due to brehmsstrahlung, $e^+\,e^-$ pair production, nuclear interactions. 
For muon transit through ice at energies in the 30~TeV to 2~PeV range, 
we follow~\cite{Barger:2011em} and take 
$a= 0.28$~TeV/km and $b = 0.28$/km  (and so $a/b = 1$~TeV. ).
Although the true energy losses are known to be stochastic rather than continuous, the average values characterized
by $a$ and $b$ parameters which we use in the continuous loss formula above are quite accurate~\cite{Gaisser:1994yf}.
 The solution to Eq.~\rf{Emuon loss} is 
 \beq{Emuon loss2}
 E_\mu (\ell) +\fracab = \left[E_\mu (0) + \fracab \right] \,
 e^{-b\ell} \, .
\eeq
The deposited energy from muon losses over the detector distance $\ell$ is then 
\beq{Emudep}
\Emudep= E_\mu(0)-E_\mu(\ell) =  \left[E_\mu(0) +\fracab \right]\, (1-e^{-b\ell}) \,.
\eeq
The number of events due to an incident muon-neutrino of energy $E_\nu$, with deposited energy $\Emudep$, is
\bea{Emudep2}
\frac{d^2 N}{d\Emudep\ dE_\numu} &=& \left( \Delta\Omega \ T \right) \
\left( A_{\rm eff}^{\numu}(E_\nu)\; \Phi_{\nu_\mu} (E_\nu) \right) 
	\int_{\Emudep}^{E_\nu} dE_\mu(0)\,\left[ \frac{1}{\sigma_{\rm
              CC} } \frac{ d\sigma_{\rm CC} }{dE_\mu(0) } (E_\mu(0),E_\nu) \right] \nonumber\\
	&& \times
	\int_{\ell_{\min}}^{\ell_{\max}} \frac{d\ell}{L} \ \delta \left\{ \Emudep - \left[ \left(E_\mu (0)+\fracab\right)\left(1-e^{-b\ell} \right) \right] \,\right\}  \,.
\eea
The length integral here averages the distance traveled by the muon in the detector,
and so $L=\ell_{\max}-\ell_{\min}$.
We take $\ell_{\min} = 300$~m so that an identifiable track is produced~\cite{GonzalezGarcia:2006na}
and take $\ell_{\max}$ equal to the IceCube detector size of 1~km.
(For future use, we note that these choices imply the values
$(1-e^{-b\ell_{\rm max}})=0.244$, and $(1-e^{-b\ell_{min}})=0.081$.)

In fact the deposited energy includes a hadronic contribution.   We define $\Edep= \Emudep + \Ehad$,
and turn to the commonly-used $y$-distribution notation for simplicity.  One defines $y\equiv \Ehad/E_\nu$.  
Then it follows that $E_\mu (0) = (1-y)\, E_\nu$, and 
% that 
% $ \int_{\Emudep}^{E_\nu} dE_\mu(0)\,\left[ \frac{1}{\sigma_\mu^{CC} } \frac{ d\sigma_\mu^{CC} }{dE_\mu(0) } (E_\mu(0),E_\nu) \right] \ 
% = \int_0^{\left(1-\frac{\Emudep}{E_\nu}\right)} \,dy\,\left[ \frac{1}{\sigma_\mu^{CC} } \frac{ d\sigma_\mu^{CC} }{dy } (y,E_\nu) \right] $\,.
one has $\Ehad= y E_\nu$, and 
\beq{Edep(Enu,ell)}
\Edep = \left[ (1-y)\,E_\nu + \fracab \right]\, \left(1-e^{-b\ell} \right) + y\,E_\nu\,.
\eeq
We arrive at
\bea{Edep}
\frac{d^2 N}{d\Edep\ dE_\numu} &=& \left( \Delta\Omega \ T \right) \
\left( A_{\rm eff}^{\numu}(E_\nu)\; \Phi_{\numu}  (E_\nu) \right)
	\int_0^1
%	{\left(1-\frac{\Emudep}{E_\nu}\right)} 
	dy\,\left[ \frac{1}{\sigma_{\rm CC} } \frac{ d\sigma_{\rm CC} }{dy } (y,E_\nu) \right] \ \nonumber\\
	&& \times
	\int_{\ell_{\min}}^{\ell_{\max}} \frac{d\ell}{L} \ 
		\delta \left\{ \Edep - \left[ \left((1-y)\,E_\nu+\fracab\right)\left(1-e^{-b\ell} \right) +yE_\nu\right] \,\right\}  \,.
\eea
The integration limits on $y$ follow from $\Edep=\Emudep+\Ehad$, where $y=1$ corresponds to pure $\Edep=\Ehad$,
and $y=0$ corresponds to pure $\Edep=\Emudep$.

Note that the $\delta$-function may be written 
\beq{delell}
\frac{\delta \left\{\ell - \ell_0\right\} } { b\,\left(\, E_\nu +a/b -\Edep \right)} \,,
\eeq
where the root $\ell_0(E_\nu,y,\Edep)$ is the effective range for the muon
\beq{ell0}
\ell_0 = \frac{1}{b}\ln\;\left( \frac{E_\nu+ a/b-y\,E_\nu}{E_\nu+a/b-\Edep} \right) 
	\approx \frac{1}{b}\,\ln\left( \frac{(1-y) }{ 1- \Edep/E_\nu} \right)\,.
\eeq
The second rendition ignore the small term $a/b\sim$~TeV.
We remark that as a check, the argument of the logarithm is always greater than one, so the log is always positive.

The $d\ell/L$~integral over the $\delta$-function is easily done analytically to yield
$ \left[(L b)\,\left(E_\nu +a/b -\Edep \right) \right]^{-1} $, while from $\ell_{\min} \le \ell_0 \le \ell_{\max}$ come  
the additional integration limits
\beq{new Emuzero limits}
y_{\min}  \; \le \; y \; \le \;  y_{\max} \,
\eeq
with 
\beq{yminmax}
y_{\min} \equiv \left( \frac{(E_\nu+a/b)\,(1-e^{+b\ell_{\max}} ) +\Edep \, e^{+b\ell_{\max}} } {E_\nu} \right)
\quad {\rm\ and\ } \quad 
y_{\max}\equiv \left( \frac{(E_\nu+a/b)\,(1-e^{+b\ell_{\min} } ) +\Edep \, e^{+b\ell_{\min} } } {E_\nu} \right) \,.
\eeq
The new extent of the $y$-range is 
$\Delta y\equiv (y_{\max}-y_{\min}) = (e^{+b\ell_{\max}}-e^{+b\ell_{\min}}) \, (E_\nu+a/b-\Edep)/E_\nu$\,.
We have
\beq{Edep2}
\frac{d^2 N}{d\Edep\ dE_\numu} = \frac{ \left( \Delta\Omega \ T \right)} {(L\,b)\,\left(E_\nu +\fracab -\Edep \right)}
	\ \left( A_{\rm eff}^{\numu}(E_\nu)\; \Phi_{\numu}  (E_\nu) \right)
	\int_{(0,\,y_{\min})}^{(1,y_{\max})}
%	{\left(1-\frac{\Emudep}{E_\nu}\right)} 
	dy\,\left[ \frac{1}{\sigma_{\rm CC} } \frac{ d\sigma_{\rm CC} }{dy } (y,E_\nu) \right] 
	  \,.
\eeq

Equation~\rf{Edep2} gives the allowed incident neutrino energies $E_\nu$ that can lead to 
the observed deposited energy $\Edep$, 
and conversely, gives the deposited energy values $\Edep$ that can result from an incident neutrino energy $E_\nu$.
The average neutrino energy giving rise to $\Edep$ is readily obtained by integrating Eq.~\rf{Edep2} over $E_\nu$ and
dividing by an appropriate $\Delta E_\nu$. Here's a parameter count:
{\it (i)}~The integration on $y$, or equivalently, choosing $\meany$, eliminates $y$;
we are left with independent $(\Edep, E_\nu, \ell)$. {\it (ii)}~Then integrating $\ell$ subject to the $\delta$-function eliminates on more variable, so we are left with two independent variables.
{\it (iii)}~The condition for the peak of $E_\nu$ versus $\Edep$ leaves just one independent variable, 
which we can take to be $E_\nu$. Thus we have $\Edep(E_\nu)$, or $\ell(E_\nu)$.
An approximate form of Eq.~\rf{Edep2} is obtained by setting $y$
equal to its average value of $\meany$, or equivalently, setting $d\sigma/dy=\sigma_0\,\delta (y-\meany)$.  
Then the final integral in Eq.~(\ref{Edep2}) equates to unity.
The trivial result is 
\beq{Edep3}
\frac{d^2 N}{d\Edep\ dE_\numu} = \frac{ \left( \Delta\Omega \ T
  \right) \ \left( A_{\rm eff}^{\numu}(E_\nu)\;  \Phi_{\numu}  (E_\nu) \right) }
	{ (Lb)\,\left( E_\nu + a/b -\Edep \right) }\,,
\eeq
Since $A_{\rm eff}^{\numu}$ is rising with $E_\nu$ only logarithmically
(see Fig.~\ref{Aeff}), and $\Phi_{\numu}$ is falling with $E_\nu$ like a power law, 
and the denominator is linearly rising  with $E_\nu$, we see that the allowed $E_\nu$ is not symmetric in its allowed region, but rather peaks at or near the lower limit.

Peak values of the exact Eq.~(\ref{Edep2}) or the approximate Eq.~(\ref{Edep3}) are given by equating the differentials
$d(\ln {\rm numerator})$ and $d(\ln{\rm denominator})$;  
so we have 
\begin{equation}
\frac{d\ln \left( A_{\rm eff}^{\numu} \Phi_{\numu} 
    \,(E_\nu) \right) }{dE_\nu}= \left(E_\nu+\fracab -\Edep \right)^{-1}
\end{equation}
 as the equation  which implicitly determines the peak value of $E_\nu$.
But $A_{\rm eff}^{\numu}(E_\nu)\;\Phi_{\numu} (E_\nu) $ is a  decreasing function of $E_\nu$, and hence its derivative
with respect to neutrino energy is negative, but equated with $(E_\nu+a/b -\Edep)^{-1}$ which is positive.
Thus the peak in the binning is backed up to the boundary value $(E_\nu)^{\rm bin\ min}$, which implies, $\ell= \ell_{\max}$. 
We get simply 
\begin{equation}
(\Edep)_{\rm peak} = \meany\,E_\nu +\left[ (1-\meany)\,E_\nu
  +\fracab\right]\,(1-e^{-b\ell_{\max}} )\,,
\label {(Edep)_{peak}1}
\end{equation}
with $E_\nu= (E_\nu)^{\rm bin\ min}$. Typically, $b\ell_{\max}$ is
${\cal O} (0.3)$ and $a/b$ is an ignorable ${\cal O}({\rm
  TeV})$, so accepting a 15\% error in the (bracketed) second term on
the right-hand side, we get finally
\beq{(Edep)_{peak}2}
(\Edep)_{\rm peak} = \left[ \meany + (1-\meany)\, b\,\ell_{\max}\right] \,(E_\nu)^{\rm bin\ min} \,;
\eeq

With $\meany\sim 0.20-0.40$,  one gets $(\Edep)_{\rm peak} \sim \,E_\nu/2$, with 
roughly half of the deposited energy arising from the hadronic deposition (the first term in Eq.~\rf{(Edep)_{peak}2}),
and half arising from the muonic deposition (the second term in Eq.~\rf{(Edep)_{peak}2}).
The two terms on the right-hand side contribute equally at $\meany=
b\,\ell_{\max} /( 1+ b\,\ell_{\max})\sim 0.22$.
Inverting Eq.~\rf{(Edep)_{peak}2} is trivial; we have 
\beq{(Edep)_{peak}3}
\left( \frac{E_\nu}{ \Edep } \right)_{\rm peak} = \left( \frac{(E_\nu)^{\rm bin\ min} }{ \Edep } \right) = \left[ \meany + (1-\meany)\, b\,\ell_{\max}\right] ^{-1} \,.
\eeq

The width at half-maximum is given by substituting the value of $\Edep$ given in Eq.~\rf{(Edep)_{peak}2} into 
Eq.~\rf{Edep3}, setting $E_\nu$ in Eq.~\rf{Edep3} equal to $E_\nu+\Gamma$, and setting the entire value 
equal to $\half$ of the peak value,  \ie,
\beq{Ewidth}
\frac{E_\nu+\Gamma}{E_\nu}= 2\,\frac{A_{\rm eff}\,\Phi_{\nu_\mu}
  (E_\nu+\Gamma)}{ A_{\rm eff}\,\Phi_{\nu_\mu}\,(E_\nu)} \,.
\eeq
For example, if $A_{\rm  eff}\,\Phi_{\nu_\mu} (E_\nu)$ behaves as a power law with index $E_\nu^{-\beta}$, then 
$\Gamma = (2^{1/(1+\beta)}-1)\,E_\nu$;
for $\beta=3.5$, we get $\Gamma = 0.165\,E_\nu$, 
 and for $\beta=2.0$, we get $\Gamma = 0.260\,E_\nu$.

For $\nu$ scattering off of a valence quark (antiquark), helicity considerations give $\meany=$1/2 (1/4),
while for $\bar \nu$ scattering, $\meany$ has the opposite values, 1/4 (1/2).
Since the target is a combination of quarks and antiquarks, one might expect $\meany$ to be bounded by 
$0.5 > \meany > 0.25$.  In fact, when sea quarks and antiquarks dominate over valence quarks, one might expect the 
averaged value of $(0.50+0.25)/2 = 0.375$ for $\meany$, for both quarks and antiquarks.  
However, for the sea quarks $\meany$ is determined in part by nontrivial integration limits
$y_{\min} > 0$, and $y_{\max} < 1$, and in part by the splitting functions of the partons.  
As a consequence, $\meany$ may and does dip below 0.25 at energies $\agt$~ PeV\@.  
In~\cite{Gandhi:1995tf}  it is shown that $\meany$  is 0.4, 0.32, 0.30, 0.27, and 0.22 at 
$E_\nu =$~30~TeV, 100~TeV, 200~TeV, 2~PeV, EeV ($10^3$~PeV), respectively, and asymptotes at 0.20 above an EeV.
The charged-current cross sections $\sigma_{CC}^\nu$ and
$\sigma_{CC}^{\bar \nu}$ retain some memory of the valence quarks at 30~TeV, 
but are nearly equal above 100~TeV.
In the weighting for $\meany$, we have taken this into account.
Values are given in Table~\ref{table:seven}.
These values, which validate the issue of energy transfer due to neutrino scattering raised in the previous section,
correspond to a fractional energy $E_\mu(0)/E_\nu$ of $1-\meany =$~0.6, 0.65, 0.70, 0.75, and 0.8, 
respectively. \footnote{
At and above $E_\nu =$~a PeV, the fractional energy of the muon $E_\mu (0)$ rises  slowly as 
$1-\meany = 0.75+0.01 \log (E_\nu/{\rm PeV}) $, 
but these energies are beyond the concerns of the present work.} 
For the 30-100~TeV data set, we set the fractional energy $E_\mu(0)/E_\nu$ to 0.62, 
for the 100-200~TeV data set, to 0.67, and for the 200~TeV-2~PeV data set, to 0.72.

\begin{table}[h]
 \caption{ For $A_{\rm eff}\,\Phi_{\numu}\propto E_\nu^{-\beta}$, average $y$, peak values for $( \Edep/E_\nu)_{\rm peak}$ 
 and $(E_\nu/\Edep)_{\rm peak}$.  The WHM ($\Gamma/E_\nu$)  for fixed $\Edep$ is $2^{1/(1+\beta)} -1$, as explained in this Appendix.}
% at half maximum 
\begin{tabular}{||c|c|c|c||}
\hline\hline
 ~~~~~~~~~~$E_\nu$~~~~~~~~~ & ~~~~~~~~~$\meany$ ~~~~~~~~~~ & ~~~~~~~~~~$( \Edep/E_\nu)_{\rm peak}$~~~~~~~~~~ 
 	&  ~~~~~~~~~~$(E_\nu/\Edep)_{\rm peak}$~~~~~~~~~~~~ \\
 \hline\hline
 10~TeV   & 0.40 & 0.57 & 1.76   \\ \hline
 100~TeV & 0.32 & 0.51 & 1.96  \\ \hline 
 200~TeV & 0.30 & 0.50 & 2.02  \\ \hline 
 PeV         & 0.27 & 0.47 & 2.11  \\ \hline
 10~PeV	& 0.25 & 0.46 & 2.17	  \\ \hline
 EeV         & 0.22 & 0.44 & 2.28   \\ \hline 
 ZeV         & 0.20 & 0.42  & 2.36  \\ 
\hline\hline 
\end{tabular}
\label{table:seven}
\end{table}

\section{Muon self-veto}
\label{sec:App2}

The upper limit of the prompt flux cannot be directly used to predict
the number of events as in (\ref{eq:ntotal}).  The veto analysis
technique implemented to avoid contamination of the sample by
atmospheric neutrinos significantly reduces the \emph{effective} flux
that will produce prompt related events at IceCube. The prompt
flux~\cite{Francis-Logan} has to be properly reduced, as described
in~\cite{Halzen:2016pwl,Halzen:2016thi}, in order to use it with the
usual effective areas. Although the north hemisphere flux remains
unaltered by the veto, the southern flux is notably reduced, as can be
seen in Fig. 8 of~\cite{Halzen:2016pwl}. In this Appendix we 
stress the relevance of taking into account the veto technique. 

\begin{figure}
\postscript{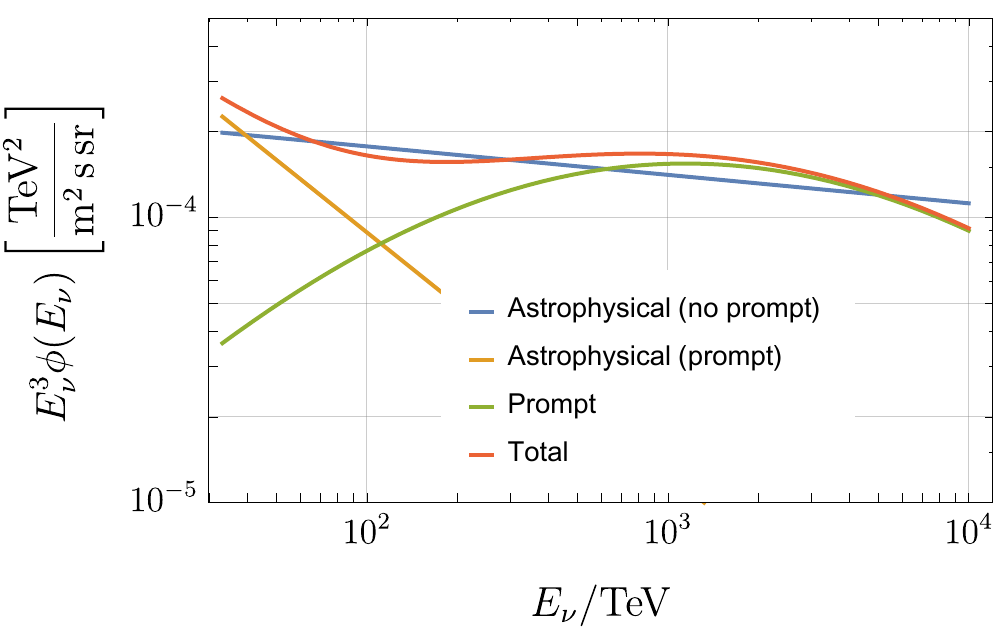}{0.5}
\caption{Prompt (no veto), astrophysical and prompt + astrophysical total fluxes for the single-exponential (showers only).}
\label{prompt_se_sh_noveto}
\end{figure}

\begin{figure}
\postscript{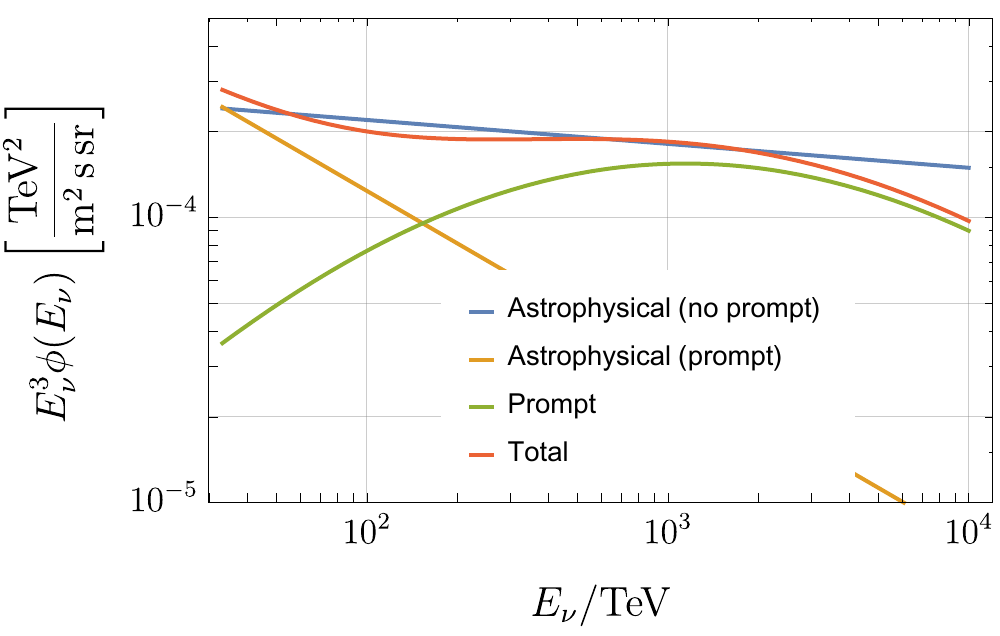}{0.5}
\caption{Prompt (no veto), astrophysical and prompt + astrophysical total fluxes for the single-exponential (showers and tracks).}
\label{prompt_se_shtr_noveto}
\end{figure}

\begin{figure}
\postscript{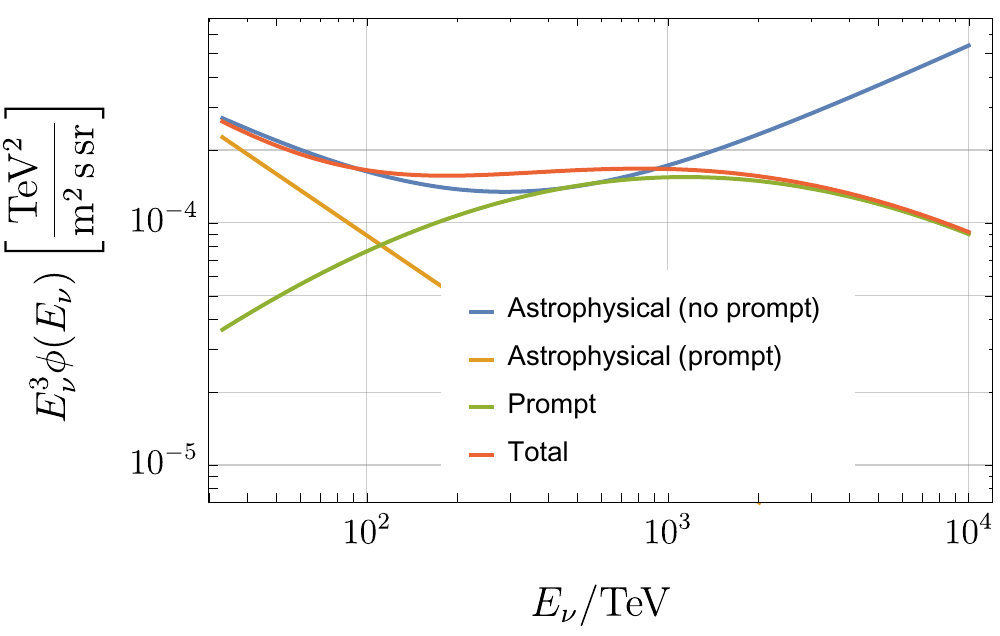}{0.5}
\caption{Prompt (no veto), astrophysical and prompt + astrophysical total fluxes for the double-exponential (showers only).}
\label{prompt_de_sh_noveto}
\end{figure}

\begin{figure}
\postscript{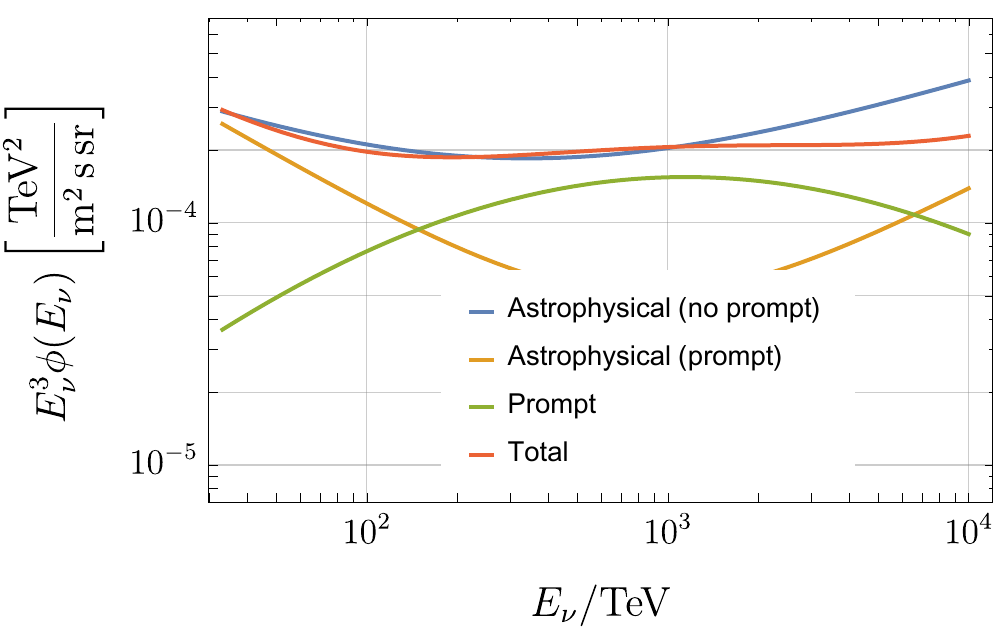}{0.5}
\caption{Prompt (no veto), astrophysical and prompt + astrophysical total fluxes for the double-exponential (showers and tracks).}
\label{prompt_de_shtr_noveto}
\end{figure}

In Figs.~\ref{prompt_se_sh_noveto}-\ref{prompt_de_shtr_noveto} we show
the results of fitting the astrophysical flux with the total
contribution of the prompt flux in the southern hemisphere, instead of
the veto reduced one. There is not room to accommodate, in any case,
an astrophysical flux through all the energy spectrum. The prompt flux
reduction at low energies, where the statistics are higher, imposes
that the astrophysical flux has to account for all the events in that
zone. But the increase of the prompt flux at higher energies makes it
impossible for the astrophysical flux to have soft exponents. Only in
Fig.~\ref{prompt_de_shtr_noveto} does the astrophysical flux dominates
again at the highest energies. But it is also unlikely that the prompt
flux could account for the highest energy events, according to its
small value.

\end{document}